\RequirePackage{fix-cm}
\documentclass{svjour3} 

\usepackage{lscape}
\usepackage{soul}
\usepackage{booktabs}
\usepackage{graphicx}
\usepackage{amssymb}
\usepackage{marginnote}
\usepackage{bbding}
\usepackage{amsmath}
\usepackage{adjustbox}
\usepackage{caption}
\usepackage{graphics}
\usepackage{float}
\usepackage{url}
\usepackage{multirow}
\usepackage{makecell}
\usepackage{longtable}
\usepackage{adjustbox}
\usepackage{rotating}
\usepackage{enumitem}
\usepackage[normalem]{ulem}
\usepackage[round]{natbib}

\makeatletter
\def\cl@chapter{\@elt {theorem}}
\makeatother

\usepackage{cleveref}[2012/02/15]
\usepackage[flushleft]{threeparttable}
\usepackage[utf8]{inputenc}
\usepackage[table]{xcolor}
\usepackage{todonotes}
\usepackage{tikz}
\usetikzlibrary{positioning, shapes.geometric}

\usepackage{tabularx}
\usepackage{ragged2e} 
\newcolumntype{L}{>{\RaggedRight}X} 
\newcommand{\smallcheckmark}{\checkmark\kern-1.1ex\raisebox{.7ex}{\rotatebox[origin=c]{125}{--}}}

\journalname{Empirical Software Engineering}

\begin{document}



\title{Does class size matter?}
\subtitle{An in-depth assessment of the effect of class size in software defect prediction}

\titlerunning{Does class size matter} 
\author{Amjed Tahir \\ Kwabena E. Bennin \\ Xun Xiao \\ Stephen G. MacDonell }

\institute{Amjed Tahir and Xun Xiao  \at
              Massey University \\
             New Zealand
              \\ \email{\{a.tahir; x.xiao\}@massey.ac.nz}           
           \and
           Kwabena E. Bennin \at
              Wageningen University \& Research, Wageningen \\
             The Netherlands
             \\
             \email{kwabena.bennin@wur.nl}
                 \and
           Stephen G. MacDonell \at
              Auckland University of Technology and University of Otago \\
              New Zealand
              \\
              \email{stephen.macdonell@otago.ac.nz}
}

\authorrunning{Tahir et al.} 

\date{Received: date }


\maketitle

\begin{abstract}
In the past 20 years, defect prediction studies have generally acknowledged the effect of class size on software prediction performance. To quantify the relationship between object-oriented (OO) metrics and defects, modelling has to take into account the direct, and potentially indirect, effects of class size on defects.
However, some studies have shown that size cannot be simply controlled or ignored, when building prediction models. As such, there remains a question whether, and when, to control for class size. 
This study provides a new in-depth examination of the impact of class size on the relationship between OO metrics and software defects or defect-proneness. We assess the impact of class size on the number of defects and defect-proneness in software systems by employing a regression-based \emph{mediation} (with bootstrapping) and \emph{moderation} analysis to investigate the direct and indirect effect of class size in \emph{count} and \emph{binary} defect prediction.
Our results show that the size effect is not always significant for all metrics. Of the seven OO metrics we investigated, size consistently has significant mediation impact only on the relationship between Coupling Between Objects (CBO) and defects/defect-proneness, and a potential moderation impact on the relationship between Fan-out and defects/defect-proneness. Other metrics show mixed results, in that they are significant for some systems but not for others. 

Based on our results we make three recommendations. One, we encourage researchers and practitioners to examine the impact of class size for the specific data they have in hand and through the use of the proposed statistical mediation/moderation procedures. 
Two, we encourage empirical studies to investigate the indirect effect of possible additional variables in their models when relevant. Three, the statistical procedures adopted in this study could be used in other empirical software engineering research to investigate the influence of potential mediators/moderators.

\end{abstract}


\section{Introduction}
\label{sec:Introduction}
Software defect prediction aims to identify potentially defective software modules which in turn aids in effective allocation and prioritisation of scarce testing resources. Constructing defect prediction models is dependent on the availability and use of historic software project data. Metrics extracted from software components have been used for predicting either the number of defects or defect-proneness in numerous software defect prediction studies \citep{d2012evaluating,Jureczko2010}. The performance and stability of prediction models have been extensively assessed and discussed in prior studies \citep{ghotra2015revisiting}.
Among the metrics often considered is class size, an attribute that is generally thought to strongly influence the relationship between all other metric features and defects. Conventionally, it is assumed that larger classes are more likely to be `defective' or be `more defective' than smaller classes \citep{mende2009revisiting}.

The jury is still out on the true impact of class size on the relationship between OO metrics and the defect-proneness of software components. Empirical evaluation of the indirect effect of class size on the relationship between software characteristics and defects has been undertaken by few researchers.
Prior studies \citep{Benlarbi2001,Zhou2014,Gil2017OnValidity} have evaluated the effect of class size on binary models in predicting defect-proneness (i.e. if a file or module is buggy or not). Notwithstanding, the conclusions of these prior studies do not apply to other defect prediction models, specifically count models built to predict the number of defects \citep{Fenton1999AModels,zhang2009investigation}. 
Additionally, the Type 1 error control and power of the statistical tests employed by prior studies \citep{Benlarbi2001,Zhou2014} (especially when causal-steps or Sobel tests are used) have been shown to be inferior to other statistical methods \citep{Hayes2013IntroductionApproach}. 

The need for the use of more robust statistical tests in validating the results of empirical analyses conducted in software engineering has been discussed in recent studies (e.g., \citep{Kitchenham2016RobustEngineering}). In the context of defect prediction, there have been recent calls to revisit many of the previous conclusions with the use of larger datasets \citep{majumder2020revisiting}. In the two previous studies \citep{Benlarbi2001,Zhou2014}, only a limited number of systems drawn from a single dataset (a single telecommunications framework with 174 classes in \citep{Benlarbi2001} and 6 open-source systems in \citep{Zhou2014})  have been used. To overcome such limitations, we therefore utilise a much larger set of systems obtained from two different datasets (with a total of 23 systems) in an attempt to provide a more general conclusion.

A two-part process has been followed in the past to examine whether a group of metrics might be useful in predicting defects/defect-proneness \citep{Basili1996HowSystems,olague2007empirical,zhou2006empirical}. In the first step, which can be referred to as ``model specification'' \citep{tantithamthavorn2018experience}, one must define which metrics should be included in the model. Each individual OO metric is used as an independent variable to build a univariate model against the dependent variable (being the number of defects or defect-proneness).
The main purpose of this process is to establish the validity of each of the selected metrics in predicting defects.
\cite{Benlarbi2001} and 
\cite{Zhou2014} found that, in this first part of the process, and for most OO metrics, class size has a significant indirect (confounding) effect on the validity of the studied OO metrics. The recommendation of both studies is that model specification using individual OO metrics should always control for class size \textit{before} researchers go on to the second step of the process, building multivariate prediction models.

The importance of such empirical hypothesis testing  when considering whether to control for the indirect effect of metrics in defect models has also been highlighted in the work of 
\cite{tantithamthavorn2018experience}. In that research \citep{tantithamthavorn2018experience}, the authors pointed out that defect models, besides predicting defects, are also critical in carefully examining phenomena and hypotheses of interest in an effort to derive empirical theories.
Similarly, in this study we are interested in studying if the OO metrics are indirectly influenced by class size as a variable in the model (i.e., the potential indirect effect of size). Thus, the main motivation of this paper is to provide a more complete understanding of the indirect effect of class size on the OO metrics' defect prediction ability. We reassess the effect of class size on the association between OO metrics and defects in a class using robust statistical methods through detailed \textit{mediation} and \textit{moderation} analyses on 23 projects (including both open source and industrial projects). 

Mediation and moderation statistical analysis methods examine the indirect effect of \textit{additional} variables on the relationship between a predictor and response variables in a model. In short, \textit{mediation} analysis tests whether the effect of the predictor (X) on the outcome (Y) operates through a third variable called the mediator (M). \textit{Moderation} analysis tests whether a moderator variable (Z) affects the direction or the strength of the relationship between the predictor (X) and the outcome (Y) variables. In other words, mediation investigates the \textit{conditional effect} of a third variable whereas moderation investigates the \textit{interaction effect} of a third variable in the model.

The main contribution of this study is an in-depth understanding of the true indirect effect (through \emph{mediation} and \emph{moderation} analysis) of class size in defect prediction models across both count and binary models. We also provide an easy-to-follow statistical procedure to assess the \emph{mediation} and \emph{moderation} effects before building prediction models.
The paper extends our early work \citep{tahir2018revisiting} in that it examines not only the effect of size in count models (i.e., the number of defects), but also in binary models (i.e., defect-proneness), and does so on a larger number of systems (23 systems in total). 
The paper addresses the following two main research questions:

\begin{quote}
   \textbf{RQ1:} does class size have an effect on the relationships between OO metrics and the number of defects?
  
  \textbf{RQ2:} does class size have an effect on the relationships between OO metrics and defect-proneness?
 
\end{quote}

\noindent RQs 1 and 2 can be then divided into two sub-questions each: 

\begin{itemize}
\item[]\textbf{RQ1.1:} does class size have a \textit{ mediation} effect on the relationships between OO metrics and the number of defects?
\item[]\textbf{RQ1.2:} does class size have a \textit{ mediation} effect on the relationships between OO metrics and defect-proneness?

\item[]\textbf{RQ2.1:} does class size have a \textit{ moderation} effect on the relationships between OO metrics and the number of defects?
\item[]\textbf{RQ2.2:} does class size have a \textit{ moderation} effect on the relationships between OO metrics and defect-proneness?

\end{itemize}

\noindent We test the following four Null Hypotheses:
\begin{itemize}
\item[] \textbf{H01.1}: class size has no \textit{mediation} effect on the relationship between OO metrics and the number of defects.
\item[] \textbf{H01.2}: class size has no \textit{mediation} effect on the relationship between OO metrics and defect-proneness.

\item[] \textbf{H02.1}: class size has no \textit{moderation} effect on the relationship between OO metrics and the number of defects.
\item[] \textbf{H02.2}: class size has no \textit{moderation} effect on the relationship between OO metrics and the defect-proneness.

\end{itemize}

Our empirical and statistical analyses show that the evidence regarding the \emph{mediation} effect of size is inconsistent across different systems. Only the effect of size on the relationship between CBO and the number of defects or defect-proneness can be considered to be significant across most systems. Size appears to have a stronger \emph{moderation} effect on the relationship between Fan-out and defects. To explain this inconsistency of results that are obtained from different systems, we conducted further analysis to explore the similarity between different systems in our dataset and then compare the similarity outcomes with the mediation and moderation analysis results obtained from these systems. This analysis reveals that data collected from different systems exhibit rather different patterns, and that systems that are highly correlated share the same mediation outcomes, but differ in their moderation outcomes. 

We encourage software teams to consider the indirect effect of class size during defect prioritisation activities, by applying the bootstrapping mediation and moderation analysis methods that have been used in this study. If size is found to have a strong effect, then statistical methods to control that effect should be used.
The statistical mediation and moderation analysis method employed here could also have wider application in other software engineering areas such as in cost and effort estimation and change prediction.

The remainder of this paper is organised as follows: Section \ref{sec:background} presents related work in defect prediction with a focus on analysing the relationship between OO metrics, size and prediction performance. The research methodology is presented in Section \ref{sec:settings}. We present the statistical analysis and results in Section \ref{sec:results}, followed by a  discussion of the practical implications of the results. Potential threats to validity are discussed in Section \ref{sec:threats} and we conclude the study in Section \ref{sec:conclusion}.

\section{Related Work}
\label{sec:background}
Defect prediction for software components has been extensively investigated to date. Previous studies have used various metrics extracted from source code or design artifacts to predict defects \citep{Basili1995,Zimmermann2007,schroter2006predicting}. Other studies have used change metrics (e.g., number of changes, commit history, dependencies, code churn) as predictors of defects \citep{Hassan2009PredictingChanges,Ying2004,zimmermann2008predicting}.
\cite{gyimothy2005empirical} found LOC to be a good low-consistency predictor of defects, as it is easier to collect compared to many other metrics. 
\cite{Fenton1999AModels}, 
\cite{Hall2011AEngineering}, 
\cite{Shepperd2014ResearcherPrediction} and 
\cite{DAmbros2012EvaluatingComparison} all provide comprehensive overviews of defect prediction models, their structure and validity. Research in defect prediction has resulted in models for both binary/classification (e.g., \cite{song2011general}) and count (e.g., \cite{osman2018impact}) prediction.

To improve prediction performance, a range of factors have been studied, among them feature and model selection \citep{catal2009investigating,hayakawa2021novel} and class imbalance \citep{bennin2018relative,bennin2017mahakil}. Additionally, cross-project defect prediction proposed to aid software practitioners who need to test new software projects with no historical data has also been studied \citep{turhan2009relative}. Several efforts \citep{kamei2013large,kamei2016studying,pascarella2019fine} on building prediction models for identifying defect-inducing changes, known as Just-In-Time defect prediction, have been investigated.
The relationships between OO metrics and defect-proneness, in particular, have been investigated by prior studies \citep{Benlarbi2001,Zhou2014}. 
\cite{Benlarbi2001} empirically investigated the relationships between OO metrics and class defect-proneness. The authors built two logistic regression models for each OO metric considered with LOC (as the size measure) included in one of the models. To measure the magnitudes of the associations between the OO metric and defect-proneness with and without controlling for size, two odds ratios were computed for the two models. A comparison between the two odds ratios with a large percentage difference suggested that class size had a confounding effect. The results of the study indicated that the associations between most of the metrics investigated (namely, Weighted Methods per Class (WMC), Response for Class (RFC) and Coupling Between Objects (CBO)) and defect-proneness were significant before controlling for size but were not so after controlling for size. The authors thus recommended that defect prediction models should always control for size.

The approach adopted by 
\cite{Benlarbi2001} mirrors the Baron and Kenny Causal-steps (joint significance) test \citep{Hayes2009BeyondMillennium}. While simple and widely used, the causal-steps test is known to have less power than other available mediation analysis approaches \citep{Fritz2007RequiredEffect,MacKinnon2002AEffects.}. Additionally, this approach is not based on a quantification of the intervening indirect effect, but examines a series of links in a causal chain between variables (e.g., $X\rightarrow M \rightarrow Y$). A similar approach was used by 
\cite{Gil2017OnValidity}, which also followed a causal-steps approach to determine the effect of size on a number of OO metrics. However, their study found that, when controlling for size, metrics may lose their predictive power. 
The findings of El Emam et al. have been argued by other works. 
\cite{Evanco2003Comments} argued that introducing class size as an additional independent variable in a defect prediction model can result in misspecified models that lack internal consistency.
Another study reported that complexity metrics are found to have a strong correlation with LOC, and therefore the more complexity metrics provide no further information that could not be measured simply with LOC \citep{herraiz2010beyond}.

More recently, 
\cite{Zhou2014} conducted a study on the effect of class size on the association between OO metrics and defect-proneness. In contrast to the study by 
\cite{Benlarbi2001}, the authors employed the Sobel test \citep{Sobel1982AsymptoticModels}, which examines the significance of the mediation effect by calculating the standard error of $X Y$, evaluating the null hypothesis that the true effect of a mediating variable is zero. Their results showed that size has a mediation effect on the relationship between OO metrics and defect-proneness. 
Positively, the Sobel test is known to perform better (in terms of power) than the Causal-steps approach \citep{Fritz2007RequiredEffect}; however, there are a number of issues with this test, particularly its weakness due to the assumption of normality of sampling. In most cases the sampling distribution of $X\rightarrow Y$ tends to be asymmetric, with high skewness and kurtosis \citep{Stinet1990,MacKinnon2002AEffects.}. In our analysis of one of the two datasets used by 
\cite{Zhou2014} (originally generated in \citep{DAmbros2012EvaluatingComparison}), we found that the most metrics data do not follow a normal distribution (see Section \ref{sec:datasets}). In addition, the Sobel test has also been shown to score low in power and Type I error control \citep{MacKinnon2002AEffects.}.
Note that all previous studies have investigated the size effect from only one perspective, the indirect effect of size; that is, where size is a third variable and that the relationship between the metrics and defect-proneness is transmitted through this variable. As such, size is in this case a \textit{mediator} variable, and such a variable is analysed using a mediation analysis technique. 

The known issues with both techniques (i.e., the Causal-steps and Sobel tests), combined with a level of conclusion instability across different studies, led us to seek alternative approaches to investigate the true indirect effect of size. In this work we assess whether class size has a mediation effect on the relationships between OO metrics and defects (in count models) using robust statistical analysis.
In addition, we assess another size effect phenomenon known as the \textit{moderation} effect - we study whether class size moderates (i.e., affects the size or strength of) the relationship between OO metrics and the number of defects. We are not aware of any other study that has investigated the moderation effect of class size in defect prediction models. As such our results should inform the design of future models that may - or may not - employ class size as one of the predictors of defects.

\section{Research Setting and Empirical Analysis}
\label{sec:settings}

\subsection{Estimating the Indirect Effect of Variables}
\label{sec:estindirecteff}
This work aims to estimate either mediation or moderation effects of size, assuming that size might have one of the effects on the relationship between OO metrics and defects in a class. Note that the previous studies in this space (e.g., \cite{Benlarbi2001} and \cite{Zhou2014}) have investigated the \textit{mediation} effect of size in defect prediction models. This is what has been referred to as the ``confounding effect''. While it is important to confirm whether size has a mediation effect in defect prediction models, we also believe that it is important to investigate the potential \textit{moderation} effect of size (that is, the impact of size on the \textit{strength} of the relationship between OO metrics and the number of defects). Understanding the nature of the indirect effect of size is quite important especially when selecting methods to control for such an effect (i.e., methods to control for a mediator could be different than those used to control for moderators).

\begin{figure}[h]
\centering
	\caption{Path diagram illustrating a simple model for the relationship between OO metrics and defects}
	\includegraphics[width=0.80\linewidth]
    {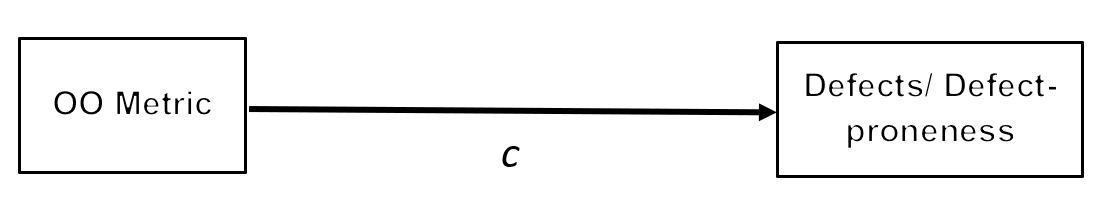}
 	\label{fig:pathDiagram_normal}
\end{figure}

\begin{figure}[h]
\centering
	\caption{Path diagram illustrating a simple \emph{mediation} effect of size on the relationships between OO metrics and defects}
	\includegraphics[width=0.80\linewidth]
        {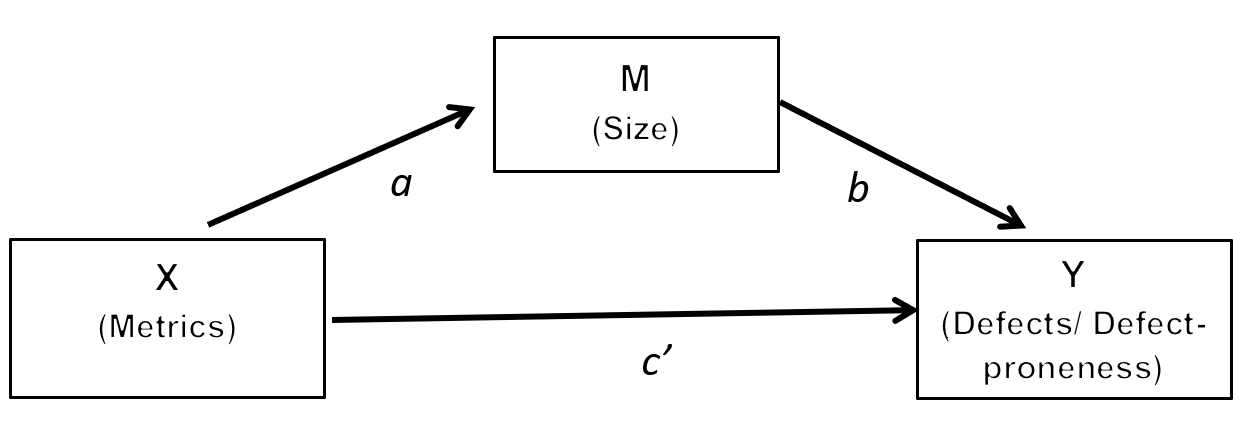}
 	\label{fig:pathDiagram_mediation}
\end{figure}

Given an independent variable $X$ and a dependent variable $Y$, the effect of $X$ on $Y$ may be transmitted through a third intervening (mediating) variable $M$.  That is, $X$ affects $Y$ because $X$ affects $M$ and $M$ in turns affects $Y$ (i.e.,  $X \rightarrow Y$ is the result of the indirect relationship  $X \rightarrow M \rightarrow Y$). Fig. \ref{fig:pathDiagram_normal} depicts a path diagram showing a direct relationship between an independent variable (IV, an OO metric) and a dependent variable (DV, number of defects or defect-proneness). Fig. \ref{fig:pathDiagram_mediation} depicts the possible indirect relationship between an OO metric and the  number of defects through a mediator: class size. If a mediator is controlled (i.e., held constant) and that mediator was the reason for the relationship between the independent and dependent variables (i.e., there is complete mediation), this indicates that there is no direct relationship between the independent and dependent variables.

\begin{figure}[ht]
    \centering
	\caption{A path diagram (a) and a statistical model (b) illustrating a simple \emph{moderation} effect of size on the relationships between OO metrics and defects/defect-proneness}
	\includegraphics[width=0.85\linewidth]
    {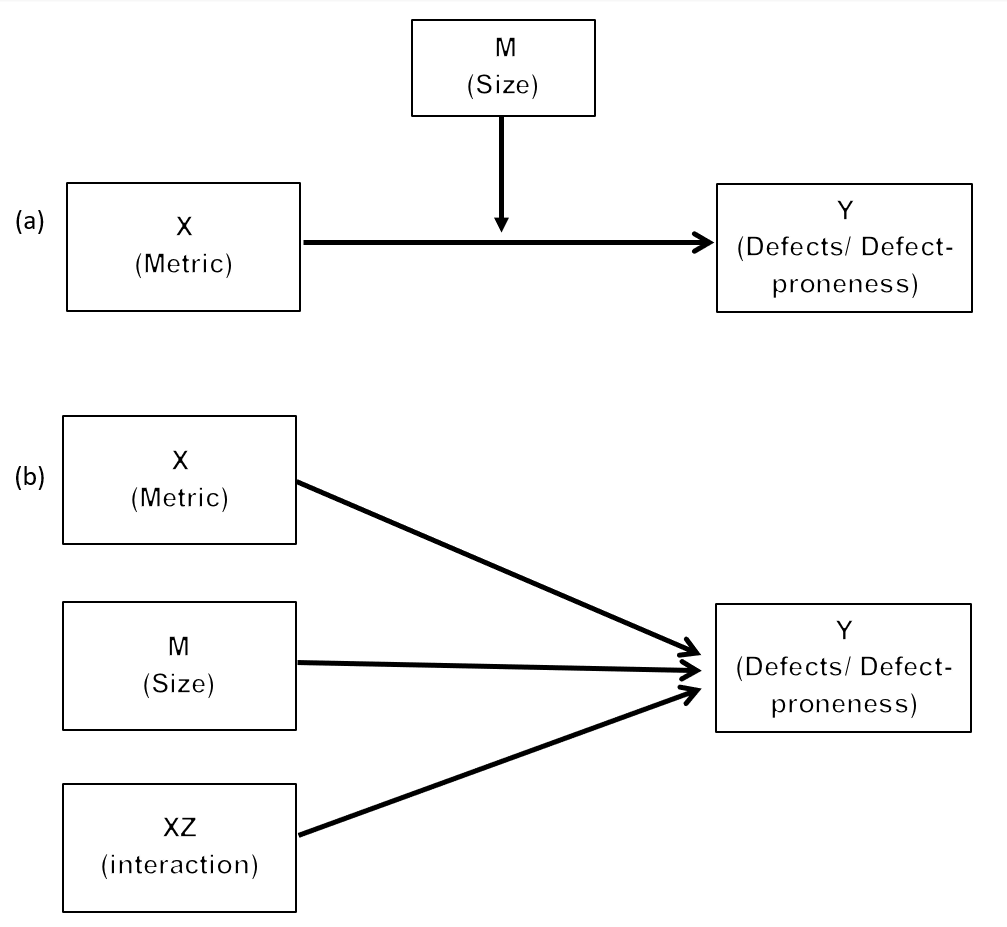}
 	\label{fig:pathDiagram_moderator}
\end{figure}

On the other hand, a \textit{moderator} is a variable that impacts the power of the relationship between the independent and dependent variables. It can affect the sign or the strength/magnitude of the relationship between those two variables.
Fig. \ref{fig:pathDiagram_moderator} shows the potential moderation effect of size on the relationship between an OO metric and defects. The relationship between  $X$ (independent variable) and $Y$ (dependent variable) is said to be moderated if its size or direction depends on a third variable $M$ \citep{Hayes2013IntroductionApproach}. In other words, a moderator is a third variable that affects the zero-ordered correlation between the independent and dependent variables.
Unlike mediation, which affects the extent that it accounts for the relationship between the independent and dependent variables, a moderation variable does not necessarily explain the relationship between variables but rather it explains \textit{the strength of} this relationship \citep{baron1986moderator}. A moderator effect is indicated by the interaction of $X$ and $M$ in explaining the outcome $Y$.

\subsection{Methods to Detect Mediators and Moderators}
\label{sec:medmodmethods}
A number of candidate approaches can be employed to determine the true effect of a mediating variable on the relationship between two other variables, and 
\cite{MacKinnon2002AEffects.} provide a detailed comparison of several well-known methods. Of those methods, there is a strong and quite recent recommendation in the literature to use bootstrapping-based mediation tests over other tests such as the causal-steps and Sobel-based tests, since it tends to have better power and Type I error control than other approaches \citep{MacKinnon2002AEffects.,Hayes2009BeyondMillennium}. 
Bootstrapping-based methods for estimating indirect effects have been discussed in the literature for some time (e.g., \cite{Stinet1990}) and the methods have received more attention in recent years \citep{MacKinnon2004ComparisonVariables,Preacher2008AsymptoticModels.,Hayes2009BeyondMillennium}. 

For mediation analysis, we adopted a bootstrapping-based method to test for different effects multiple times using a regression-based approach. However, the approach used to estimate mediation effects for count models (i.e., where the dependent variable is a count, in our case, that is the number of defects) and binary models (the dependent variable is defect-proneness) is slightly different - this is discussed in more detail in Section \ref{sec:settings:medanalysis}. 





In moderation analysis, we study whether the effect of the increase/decrease in the independent variable $X$ does not depend on a moderator $M$.
The moderator $M$ interacts with $X$ in predicting $Y$ if the regression weight of $Y$ on $X$ varies as a function of $M$.
We measure the change in the $R^2$ value due to the interaction of the moderator variable $X\cdot M$. In case that such an interaction is significant, we then used simple slope analysis to probe the interaction of the moderator. 
Simple slope can visually portray how another variable moderates the association between the independent and dependent variable. This is done when the dependent variable is either count (number of defects) or dichotomous (defect-proneness).

\subsection{Conducting Mediation and Moderation Analysis}
\label{sec:applyingAnalysis}

\subsubsection{Mediation Analysis}
\label{sec:settings:medanalysis}


In conducting our mediation analysis, we produced a series of regression models. As explained in Section \ref{sec:estindirecteff}, the mediation effect is estimated through three levels of effect: the \textit{direct}, \textit{indirect} and \textit{total} effects (illustrated in Figs \ref{fig:pathDiagram_normal} and \ref{fig:pathDiagram_mediation}).
To investigate the \textit{indirect effect} of size on the number of defects, we first build a linear model where the dependent variable (the number of defects) is explained directly by a single OO metric (step 1). This is called the \textit{direct effect} (see Fig. \ref{fig:pathDiagram_normal}). Then we build another linear model where the mediator (transformed LOC) is explained by the independent variable (the same single OO metric) (step 2). Using Poisson regression (in case where the outcomes is a non-negative integer value), we then built a new model to explain variance in the dependent variable (number of defects) using both the independent variable and the mediator (transformed LOC) (step 3) - this is the path $ab$ in Fig. \ref{fig:pathDiagram_mediation}. The \textit{indirect effect} is then estimated. 
When the outcome variable is defect-proneness (i.e., the dependent variable is dichotomous), we use logistic regression models instead in steps 1 and 3.


The reasons for those choices are rather straightforward. Defect-proneness data follows a binary-value Bernoulli distribution, and logistic regression is the standard approach for modelling binary-value random variables with possible covariates. For the number of defects, our normality analysis (discussed in Section \ref{sec:datasets}) shows that the number of defects obtained from different datasets used in this study tend not to follow a Gaussian distribution, but in fact it is closer to a Poisson distribution. Given a sufficiently large class size with $M$ LOC with a small probability $\lambda$ of defect occurrence in each line, the Poisson limit theorem guarantees that the number of defects will follow a Poisson distribution with mean $\mu=M\lambda$ approximately.  Thus, we decided to use Poisson regression as it is widely used in modelling random counts with covariates. The detailed math formulations of these models are presented next.

To model the \textit{total effect}, and in case of a Poisson regression model where the outcome and number of defects $Y$, follow a Poisson distribution with mean $\mu$, we link the log mean with each individual OO metric $X$ as:
\begin{equation}
    \log (\mu)  = \gamma_0 + \gamma_1 X
\end{equation}

On the other hand, and in the case of logistic regression where the outcome, defect-proneness $Z$, follows a Bernoulli distribution with a probability $p$, to model the \textit{total effect} we link the log odds with each individual OO metric $X$ as:

\begin{equation}
    \log \left(\frac{p}{1-p}\right)  = \beta_0 + \beta_1 X
\end{equation}

To model the \textit{direct effect}, we use a linear regression model with the mediator $\log(M)$, i.e., log class size as the outcome and each individual OO metric $X$ as the covariate

\begin{equation}
\log(M)=\theta_0 + \theta_1 X
\end{equation}

To model the \textit{indirect effect} of size, 
we include two covariates in the model, with an individual OO metric as the independent variable and class size $M$ as the mediator. However, Poisson regression imposes a log-linear relationship between the mean of $Y$ and the covariates, which can be further formulated as:

\begin{equation}
\label{equ:poisson}
\mbox{mean}[Y]=\mu=\exp\left(\gamma_0 + \gamma_1 X + \gamma_2 M\right)
\end{equation}

Equation \ref{equ:poisson} implies that the average number of defects increases with increasing class size $M$ at an exponential rate. This contradicts the Poisson limit theorem which implies that there is a linear relationship between class size $M$ and the average number of defects $\mu$.
In fact, the class size data obtained from the datasets used in our study were far from the Gaussian distribution needed for linear models (see details in Section \ref{sec:datasets}), while the log class size fits better into a bell-shaped density.
Therefore, rather than using class size directly, we use the log class size $\log(M)$ as the mediator in both Poisson and logistic regression models, as follows:

\begin{equation}
    \log (\mu)  = \gamma'_0 + \gamma'_1 X  + \gamma'_2\log(M)
\end{equation}

\begin{equation}
    \log \left(\frac{p}{1-p}\right)  = \beta'_0 + \beta'_1 X + \beta'_2\log(M)
\end{equation}

Note that we estimate the indirect effect (at step 3) only if the results of the first two steps are found to be significant (i.e., the \textit{total} and \textit{direct} effects are both significant). If the results of all three steps are found to be significant, then we establish that data are consistent with the hypothesis that variable $M$ mediates the $X$$\rightarrow$$Y$ relationship. On the other hand, if the results of either of the first two steps were not found to be significant, then we conclude that the mediation effect of $M$ is unlikely. For defect-proneness (i.e., when $Y$ is dichotomous), we estimate the indirect effect of $ab$ only when the direct effect $c'$ is significant. Similarly, if both are significant, we then establish that $M$ mediates the $X$$\rightarrow$$Y$ relationship.

The mediation analysis was carried out using the \texttt{mediation} package for \texttt{R}\footnote{\url{https://cran.r-project.org/web/packages/mediation/}}. This package provides a mechanism to conduct mediation analysis using various standard models such as linear or generalized linear models and returns the estimates of the average causal mediation effects. The \texttt{mediate} function in \texttt{Mediation} calculates point estimates, CIs and the p-values, for the average \textit{direct}, \textit{indirect}, and \textit{total} effects.
The algorithm and formal definitions of the models used in \texttt{mediation} are explained in 
\cite{tingley2014mediation}.

In applying bootstrapping mediation analysis, we employ 5000 bootstrap resamples and set the confidence intervals to 95\% for the mediation analysis. 
\cite{tingley2014mediation} suggested to use a bootstrap with a minimum of 1000 resamples. Other studies recommended 5000 bootstrap resamples for effective resampling, as long as there is enough computation resources \citep{Preacher2008AsymptoticModels.}. We first experimented with various bootstrap resamples (i.e., using 10,000 bootstrap) to see if an increase in the resampling would in fact impact the results. We compared the results obtained from running both 5000 and 10,000 bootstraps for 3 of the systems (total of 6 analyses) for each of two datasets we used in this study. We did not observe any changes in the overall results (i.e., no significance changes in p-values or CI levels). Therefore, we decided to conduct all mediation analysis using 5000 bootstrap resamples, following the recommendation in  \cite{Preacher2008AsymptoticModels.}.



\subsubsection{Moderation Analysis}
\label{sec:settings:modanalysis}
Moderation analysis is also estimated via regression modelling by building a regression model to fit the outcome (number of defects or defect-proneness) using the independent variable (an OO metric), the moderator and the \textit{interaction term}, which tests whether the relationship between the OO metric and the outcome changes based on the moderator. The interaction term is computed as the product of the independent variable and the moderator. If the interaction term has $p<.05$, the moderation effect is then declared as significant.

Initially, we investigated those interaction terms at two levels, one for each individual metric (i.e., single predictor) and with all metrics combined (i.e., multivariate predictors). Although we investigated both approaches, the interactions between the individual metrics and the moderators, tested individually, would provide a clearer view of the true effect of the moderator.  If we build a model with all interactions put together, and that model has a significantly better fit than the model without interactions, we will likely not know for sure which interaction is independently significant and will have more impact; thus, the decision to use univariate models to investigate the moderation effect of size.

Similar to the mediation analysis explained in Section \ref{sec:settings:medanalysis},  we used Poisson and logistic regressions to model the moderation effect of size, using log class size $\log(M)$ as the moderator. 
In models where the outcome is the number of defects $Y$ and in which the size data follows a Poisson distribution with mean $\mu$, we link the log mean with the OO metric $X$, the moderator $\log(M)$ and their interaction term $X\cdot\log(M)$ as:

\begin{equation}
    \log (\mu)  = \gamma_0 + \gamma_1 X +\gamma_2 \log(M) +\gamma_3 X\cdot\log(M)
\end{equation}

For models where the outcome is defect-proneness $Z$ and size follows a Bernoulli distribution with a probability $p$, we link the log odds with the OO metric $X$, the moderator $\log(M)$ and their interaction term $X\cdot\log(M)$  as:

\begin{equation}
    \log \left(\frac{p}{1-p}\right)  = \beta_0 + \beta_1 X +\beta_2 \log(M) +\beta_3 X\cdot\log(M)
\end{equation}


\subsection{Datasets}
\label{sec:datasets}
To examine the potential size effect on defect models, we selected two publicly available open source datasets (
\cite{DAmbros2012EvaluatingComparison} (DAMB), and 
\cite{Jureczko2010} (JURE)). 
We select these datasets and their constituent systems for the following reasons: 1) they contain systems of large size which have been developed by popular open-source communities such as Eclipse and Apache, 2) the JURE dataset contain data obtained from both open-source and and proprietary software (6 industrial systems) so our data sample can be \textit{slightly} representative, 3) these datasets have been widely used in previous defect prediction studies (e.g., \citep{he2012investigation,bennin2017mahakil}) -- in particular the DAMB dataset was used in the study of 
\cite{Zhou2014}, which offers an opportunity to compare results.
We also considered other datasets such as 
\cite{zimmermann2008predicting} and the NASA metrics program data, 
but we found that those datasets do not report the same metrics as those already selected for our study (i.e., \cite{zimmermann2008predicting}) or they are too small (i.e., KC1), therefore they are not suitable for our study. Both datasets are publicly available.

The 
\cite{DAmbros2012EvaluatingComparison} dataset includes metrics data collected from five large open source projects from two well-established ecosystems: Eclipse (JDT Core, PDE UI, Mylyn and Equinox) and Apache (Lucene). 
For simplicity, we refer to this dataset as DAMB in the rest of this paper. The 
\cite{Jureczko2010} dataset contains data from open source and proprietary projects (although from relatively small systems). 
We excluded systems comprising small number of modules (fewer than 30) as we consider them too small for our analysis. 
We included 18 systems from this dataset in our study. 
This includes 11 well-known Apache projects (including Tomcat, Ant, Log4j and Xalan) and jEdit (a popular text/code editor). We refer to this dataset as JURE. 

\begin{figure*}[h]
	\centering
    \captionsetup{justification=centering}
\includegraphics[width=\linewidth]
    {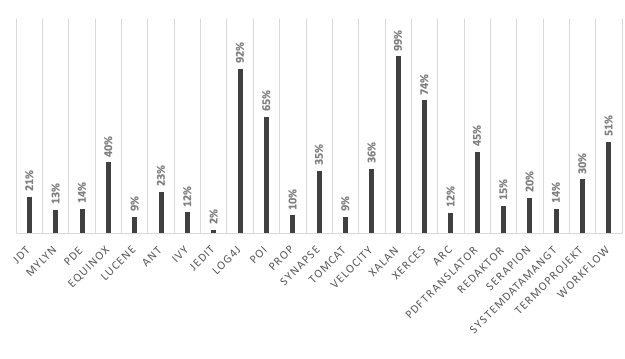}
   \caption{Percentage of defective modules (files) across all systems}
    \label{fig:Data_distribution}
\end{figure*}

We thus considered 23 systems in our study, containing a total of 11.4k files and 2.5m LOC . Note that all those systems from the DAMB and JURE datasets are Java based.
The systems we selected here contain different levels and proportions of defective modules.
General information about the systems in the datasets is shown in Table \ref{tab:datasets}. Fig. \ref{fig:Data_distribution} shows the distribution of defective modules across all systems in both datasets.
More detailed information about the selected systems, including versioning information, are provided in the original studies \citep{DAmbros2012EvaluatingComparison,Jureczko2010}. 

\begin{table}[h]
\centering
\caption{Details of the systems used in the empirical study}

\label{tab:datasets}

\begin{tabular}{lcccccc}
\toprule
\centering
Dataset & \#projects & NOC    & LOC       & \#defects & \begin{tabular}[c]{@{}c@{}}\#defective \\ classes\end{tabular}  & \begin{tabular}[c]{@{}c@{}}\%defective \\ classes\end{tabular} \\ \midrule
DAMB    & 5          & 5,238   & 639,826   & 1,740    & 843              & 16\%             \\
JURE    & 18         & 6,206   & 1,879,894 & 4,863    & 2,419          & 39\%            \\ \midrule
Total   & 23         & 11,444 & 2,519,720 & 6,603    & 3,262            & 29\%   \\ \bottomrule
\end{tabular}
\end{table}

\begin{figure}[h]
  \centering
   \resizebox{\linewidth}{!}{%

  \begin{tabular}{c @{\qquad} c }
      \includegraphics[width=\linewidth]{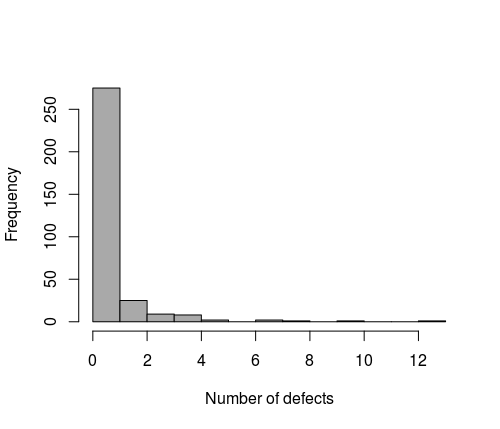} &
     \includegraphics[width=\linewidth]{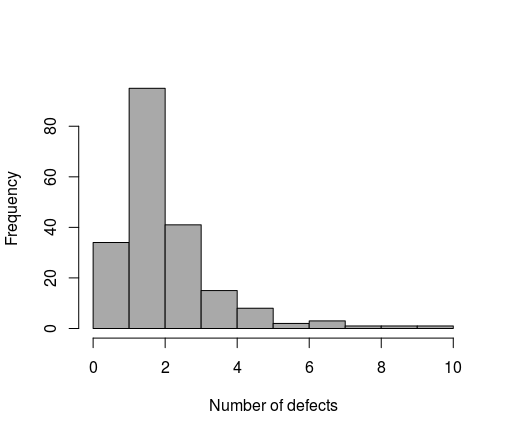} \\
    \Large	 (a) Eclipse Equinox & \Large	 (b) Apache Log4j
  \end{tabular}}
  \caption{Distribution of defects in two selected systems from DAMB and JURE datasets}
   \label{fig:defects}
\end{figure}

\begin{figure}
  \centering
   \resizebox{\linewidth}{!}{%

  \begin{tabular}{c @{\qquad} c }
         \includegraphics[width=\linewidth]{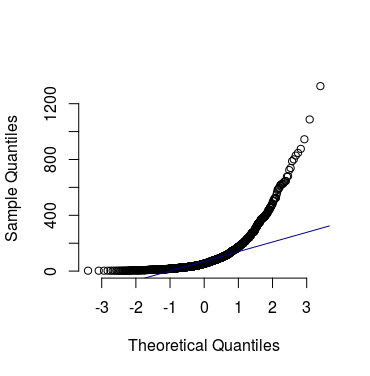}
&  \includegraphics[width=\linewidth]{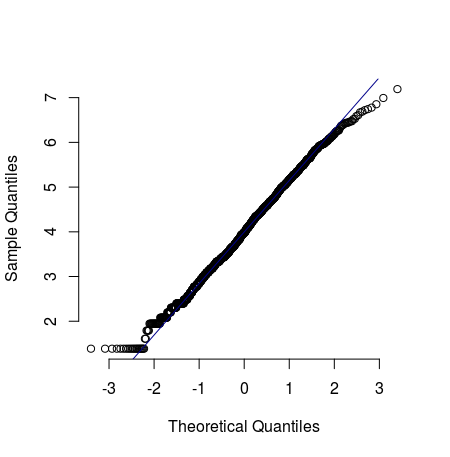}  \\
    \Large	 (a) before & \Large	 (b) after
  \end{tabular}}
\caption{Probability plots of LOC in Eclipse PDE before and after transformation}
   \label{fig:loc_PDE}
\end{figure}

\begin{figure}
  \centering
   \resizebox{\linewidth}{!}{%

  \begin{tabular}{c @{\qquad} c }
        \includegraphics[width=\linewidth]{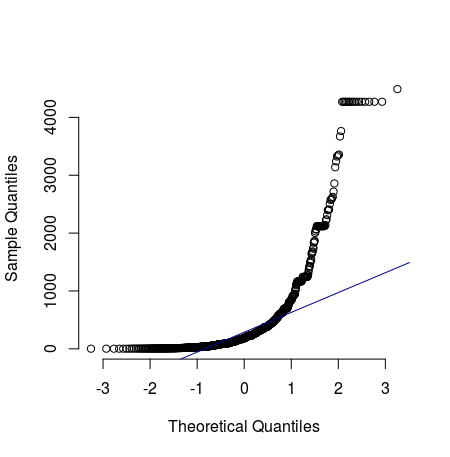}
& \includegraphics[width=\linewidth]{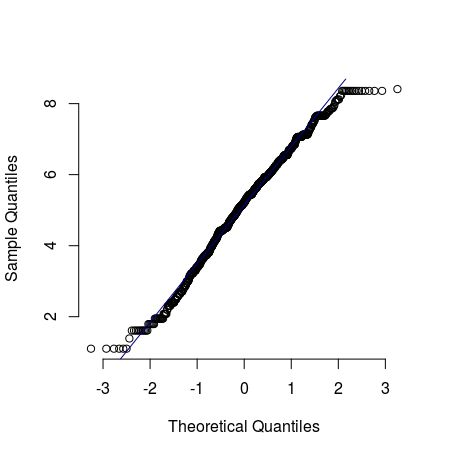}  \\
    \Large	 (a) before & \Large	 (b) after
  \end{tabular}}
\caption{Probability plots of LOC in Apache Xalan before and after transformation}
\label{fig:loc_Xalan}
\end{figure}

To inform the models' selection, we examined the normality of the various data distributions for and LOC and defects using the Shapiro-Wilk test. For the DAMB and JURE datasets, we test the null hypothesis that the data follow a normal distribution. The results of this test lead us to reject the null hypothesis that the samples come from a normal distribution. 
We provide two examples of the shape of defects from our datasets in Fig. \ref{fig:defects} (for defects in Log4J (JURE dataset) and Eclipse Equinox (DAMB dataset)). These histograms visually confirm that those systems (and other systems that exhibit a similar patter) tend to follow a distribution that is very far from Gaussian distribution; most defects are concentrated in certain classes or files, and most files contain few defects.
We also use probability plots (a graphical technique to assess if the data follows a given distribution) to plot the distribution of class size. Figs. \ref{fig:loc_PDE}(a) and \ref{fig:loc_Xalan}(a) show probability plots for LOC in Eclipse PDE (DAMB dataset) and Apache Xalan (JURE dataset), respectively. In general, classes with very prominent modes are larger than lower mode classes, while the majority of classes are under 500 LOC. Figs. \ref{fig:loc_PDE}(b) and \ref{fig:loc_Xalan}(b) show that the shape of the LOC data tends to follow a normal (or near normal) distribution after transformation (discussed in more details in Section \ref{sec:applyingAnalysis}). 

We studied the effect of class size (using the transformed LOC (i.e., $\log(LOC)$)) on the relationship between each of the following OO metrics and the number of defects/defect-proneness: RFC, WMC, CBO, DIT, LCOM, Fan-in and Fan-out as shown in Table \ref{tab:metrics}. These metrics cover several OO design quality criteria representing the internal and external complexity of the class. These include coupling, cohesion, inheritance and complexity. RFC, CBO, WMC, LCOM and DIT are part of the CK metrics suite \citep{Chidamber1994}.

\begin{table}[h]
	\caption{List of metrics used}
    \centering
	\begin{tabular}{l|l|l|l}
\toprule
\textbf{Metric}  & \textbf{Full name }            & \textbf{Metric} & \textbf{Full name}                               \\ \hline
LOC     & Lines of Code          & CBO     & Coupling between Objects                \\
RFC     & Response for Class    & DIT     & Depth of Inheritance Tree               \\
WMC     & Weighted Method Count & LCOM    & Lack of Cohesion in Methods             \\
Fan-in & \begin{tabular}[c]{@{}l@{}} No. of other classes that reference \\ the class \end{tabular}            & Fan-out & \begin{tabular}[c]{@{}l@{}}No. of classes referenced by \\ the class \end{tabular} \\ \bottomrule

\end{tabular}
	\label{tab:metrics}
\end{table}

We apply our mediation and moderation analysis procedures to seven OO metrics, LOC and defects. For each system in our datasets, the total number of tests conducted is $4\times7$ = 28 tests for both moderation and mediation analysis.  We conducted all analysis on \texttt{R}\footnote{Version 3.6.0} on a machine running on a 2.6 GHz Intel Core i7 processor, and 32 GB RAM. Each of our analysis scripts ran in under 1 hour. For reproducibility and replication, we provide all results and scripts in our online repository \citep{repository}.

\section{Results and Discussion}
\label{sec:results}



As explained in Section \ref{sec:applyingAnalysis}, we followed a three-steps approach to investigate the potential effect of size. We first examine the associations between OO metrics and the number of defects. Given that the data are not normally distributed, we use the non-parametric Kendall's tau ($\tau$) correlation coefficient, and correlation strength is interpreted using Cohen's classification (low correlation when 0 $<$ $\tau$ $\leq$ .29, medium when .30 $\leq$ $\tau$ $\leq$ .49 and high when $\tau$ $\geq$ .50). We also examine the correlation between the size metric (measured in LOC), OO metrics and the number of defects.

\subsection{The Correlation Between Size, Defects and OO Metrics}
Several previous studies used bivariate correlations and univariate logistic regressions to establish the validity of those metrics in predicting defects \citep{binkley1998validation,harrison1998coupling} or defect-proneness \citep{briand2000exploring,tang1999empirical}.
In this section, we examined the bivariate correlations between size, OO metrics and the number of defects across for each of the systems in both datasets. 
We compute Kendall's $\tau$ (Tau) correlation test between the individual metrics and defects for both datasets` (Tables \ref{tab:correlation_ken_DAMB} and \ref{tab:correlation_ken_JURE}). The correlation values for both count of defects and binary defects were mostly significant. Positive and significant correlations between the pairs (metric and defects) are observed across almost all datasets. Few insignificant correlation values (highlighted values in grey) are observed in the JURE datasets. The Fan-in metric has most of the insignificant correlation values whereas LOC, Fan-out, RFC and WMC metrics had more significant correlation values across all datasets. There were no insignificant correlation values for the DAMB datasets indicating the strong correlation between all individual metrics and defects.

\begin{table}[h]
\centering
\caption{Kendall {$\tau$} rank correlation analysis between each
individual metric and count/binary defects for DAMB
dataset. All values are significant (p $>$0.05)}
\begin{tabular}{@{}llllllll@{}}
\toprule
Metric  & Project & Defects & Defective & Metric & Project & Defects & Defective \\ \midrule
        & JDT     & 0.31    & 0.30      &        & JDT     & 0.17    & 0.16      \\
        & Equinox & 0.44    & 0.41      &        & Equinox & 0.37    & 0.36      \\
CBO     & Lucene  & 0.15    & 0.15      & Fan-in & Lucene  & 0.15    & 0.15      \\
        & Mylyn   & 0.20    & 0.20      &        & Mylyn   & 0.11    & 0.10      \\
        & PDE     & 0.18    & 0.17      &        & PDE     & 0.05    & 0.05      \\ \midrule
        & JDT     & 0.33    & 0.32      &        & JDT     & 0.24    & 0.23      \\
        & Equinox & 0.38    & 0.34      &        & Equinox & 0.45    & 0.42      \\
Fan-out & Lucene  & 0.11    & 0.11      & LCOM   & Lucene  & 0.15    & 0.15      \\
        & Mylyn   & 0.23    & 0.23      &        & Mylyn   & 0.12    & 0.11      \\
        & PDE     & 0.21    & 0.21      &        & PDE     & 0.17    & 0.17      \\ \midrule
        & JDT     & 0.32    & 0.31      &        & JDT     & 0.34    & 0.33      \\
        & Equinox & 0.41    & 0.38      &        & Equinox & 0.44    & 0.41      \\
RFC     & Lucene  & 0.12    & 0.12      & WMC    & Lucene  & 0.14    & 0.14      \\
        & Mylyn   & 0.19    & 0.19      &        & Mylyn   & 0.14    & 0.14      \\
        & PDE     & 0.22    & 0.22      &        & PDE     & 0.19    & 0.19      \\ \midrule
        & JDT     & 0.33    & 0.32      &        &         &         &           \\
        & Equinox & 0.42    & 0.40      &        &         &         &           \\
LOC     & Lucene  & 0.13    & 0.13      &        &         &         &           \\
        & Mylyn   & 0.19    & 0.18      &        &         &         &           \\
        & PDE     & 0.21    & 0.21      &        &         &         &           \\ \bottomrule
\end{tabular}
\label{tab:correlation_ken_DAMB}

\end{table}

\definecolor{Gray}{gray}{0.9}
\begin{landscape}
\begin{table}[h]
\caption{Kendall {$\tau$} rank correlation analysis between each
individual metric and count/binary defects for JURE
dataset. Non-significant correlation results (p $>$0.05) are
highlighted in gray}
\resizebox{\linewidth}{!}{%
\begin{tabular}{cllllllllllll}
\hline 
 & \multicolumn{12}{c|}{{JURE DATASETS}} \tabularnewline
\hline 
\textbf{Metric} & \textbf{Project} & \textbf{\# Defects} & \textbf{Defective} & \textbf{Project} & \textbf{\# Defects} & \textbf{Defective} & \textbf{Project} & \textbf{\# Defects} & \textbf{Defective}  \tabularnewline
\hline 
 & {Ant} & {0.29} & {0.28} & {PDFTranslator} & {0.45} & {0.55} & {Synapse} & {0.24} & {0.24} & {WorkFlow} & {\cellcolor{gray!40}-0.20} & {\cellcolor{gray!40}-0.20}\tabularnewline
 & {arc} & {0.24} & {0.24} & {poi-3.0} & {0.40} & {0.35} & {SystemData} & {0.31} & {0.32} & {Xalan} & {\cellcolor{gray!40}-0.01} & {0.07}\tabularnewline
{CBO} & {Ivy} & {0.25} & {0.25} & {Prop} & {0.14} & {0.14} & {TermoProjekt} & {0.27} & {\cellcolor{gray!40}0.25} & {Xerces} & {0.50} & {0.53} \tabularnewline
 & {jEdit} & {0.07} & {\cellcolor{gray!40}0.07} & {Redaktor} & {\cellcolor{gray!40}0.05} & {\cellcolor{gray!40}0.06} & {Tomcat} & {0.25} & {0.25} &  &  &  \tabularnewline
 & {log4j} & {0.28} & {0.13} & {Serapion} & {0.31} & {0.30} & {Velocity} & {0.26} & {0.27} &  &  & \tabularnewline
\hline 
 & {Ant} & {0.12} & {0.11} & {PDFTranslator} & {0.43} & {0.42} & {Synapse} & {0.21} & {0.20} & {WorkFlow} & {\cellcolor{gray!40}-0.27} & {\cellcolor{gray!40}-0.27} \tabularnewline
 & {arc} & {\cellcolor{gray!40}-0.11} & {\cellcolor{gray!40}-0.11} & {poi-3.0} & {0.20} & {0.13} & {SystemData} & {0.24} & {0.24} & {Xalan} & {-0.08} & {\cellcolor{gray!40}0.00}\tabularnewline
{Fan-in} & {Ivy} & {\cellcolor{gray!40}0.07} & {\cellcolor{gray!40}0.07} & {Prop} & {\cellcolor{gray!40}0.01} & {0.01} & {TermoProjekt} & {\cellcolor{gray!40}0.13} & {\cellcolor{gray!40}0.10} & {Xerces} & {0.32} & {0.41} \tabularnewline
 & {jEdit} & {\cellcolor{gray!40}0.04} & {\cellcolor{gray!40}0.04} & {Redaktor} & {\cellcolor{gray!40}-0.06} & {\cellcolor{gray!40}-0.05} & {Tomcat} & {0.16} & {0.16}\tabularnewline
 & {log4j} & {0.27} & {0.14} & {Serapion} & {0.40} & {0.40} & {Velocity} & {\cellcolor{gray!40}0.01} & {\cellcolor{gray!40}0.01} &  &  &  \tabularnewline
\hline 
 & {Ant} & {0.30} & {0.29} & {PDFTranslator} & {0.46} & {0.52} & {Synapse} & {0.21} & {0.20} & {WorkFlow} & {\cellcolor{gray!40}0.20} & {\cellcolor{gray!40}0.20} \tabularnewline
 & {arc} & {0.28} & {0.28} & {poi-3.0} & {0.40} & {0.39} & {SystemData} & {0.25} & {0.25} & {Xalan} & {\cellcolor{gray!40}0.03} & {0.07} \tabularnewline
{Fan-out} & {Ivy} & {0.17} & {0.16} & {Prop} & {0.12} & {0.12} & {TermoProjekt} & {0.39} & {0.40} & {Xerces} & {0.55} & {0.55}\tabularnewline
 & {jEdit} & {0.11} & {0.11} & {Redaktor} & {0.14} & {0.14} & {Tomcat} & {NA} & {NA} &  &  &  \tabularnewline
 & {log4j} & {0.12} & {\cellcolor{gray!40}0.01} & {Serapion} & {\cellcolor{gray!40}0.05} & {\cellcolor{gray!40}0.05} & {Velocity} & {0.31} & {0.33} &  &  &  \tabularnewline
\hline 
 & {Ant} & {0.33} & {0.33} & {PDFTranslator} & {0.57} & {0.63} & {Synapse} & {0.15} & {0.15} & {WorkFlow} & {\cellcolor{gray!40}0.08} & {\cellcolor{gray!40}0.08} \tabularnewline
 & {arc} & {0.18} & {0.18} & {poi-3.0} & {0.28} & {0.33} & {SystemData} & {0.29} & {0.27} & {Xalan} & {0.11} & {\cellcolor{gray!40}0.05} \tabularnewline
{LCOM} & {Ivy} & {0.24} & {0.24} & {Prop} & {0.12} & {0.12} & {TermoProjekt} & {0.33} & {0.33} & {Xerces} & {0.12} & {\cellcolor{gray!40}0.01} \tabularnewline
 & {jEdit} & {\cellcolor{gray!40}0.05} & {\cellcolor{gray!40}0.05} & {Redaktor} & {\cellcolor{gray!40}-0.13} & {\cellcolor{gray!40}-0.13} & {Tomcat} & {0.18} & {0.18} &  &  & \tabularnewline
 & {log4j} & {0.12} & {\cellcolor{gray!40}0.07} & {Serapion} & {0.51} & {0.50} & {Velocity} & {\cellcolor{gray!40}0.08} & {0.12} &  &  &  \tabularnewline
\hline 
 & {Ant} & {0.40} & {0.39} & {PDFTranslator} & {0.50} & {0.54} & {Synapse} & {0.34} & {0.34} & {WorkFlow} & {0.31} & {0.31} \tabularnewline
 & {arc} & {0.18} & {0.18} & {poi-3.0} & {0.42} & {0.40} & {SystemData} & {0.28} & {0.27} & {Xalan} & {0.14} & {0.09} \tabularnewline
{RFC} & {Ivy} & {0.29} & {0.29} & {Prop} & {0.16} & {0.16} & {TermoProjekt} & {0.47} & {0.48} & {Xerces} & {0.35} & {0.23} \tabularnewline
 & {jEdit} & {\cellcolor{gray!40}0.05} & {\cellcolor{gray!40}0.05} & {Redaktor} & {\cellcolor{gray!40}0.00} & {\cellcolor{gray!40}0.00} & {Tomcat} & {0.26} & {0.26} &  &  &  \tabularnewline
 & {log4j} & {0.16} & {\cellcolor{gray!40}0.05} & {Serapion} & {0.34} & {0.33} & {Velocity} & {0.23} & {0.25} &  &  &  \tabularnewline
\hline 
 & {Ant} & {0.35} & {0.34} & {PDFTranslator} & {0.48} & {0.53} & {Synapse} & {0.25} & {0.25} & {WorkFlow} & {\cellcolor{gray!40}0.19} & {\cellcolor{gray!40}0.19}\tabularnewline
 & {arc} & {0.19} & {0.19} & {poi-3.0} & {0.37} & {0.40} & {SystemData} & {0.25} & {0.24} & {Xalan} & {0.07} & {0.07} \tabularnewline
{WMC} & {Ivy} & {0.28} & {0.28} & {Prop} & {0.14} & {0.15} & {TermoProjekt} & {0.46} & {0.47} & {Xerces} & {0.23} & {0.10} \tabularnewline
 & {jEdit} & {\cellcolor{gray!40}0.05} & {\cellcolor{gray!40}0.05} & {Redaktor} & {\cellcolor{gray!40}-0.10} & {\cellcolor{gray!40}-0.10} & {Tomcat} & {0.20} & {0.20} &  &  &  \tabularnewline
 & {log4j} & {0.22} & {0.13} & {Serapion} & {0.52} & {0.50} & {Velocity} & {0.21} & {0.25} &  &  & \tabularnewline
\hline 
 & {Ant} & {0.39} & {0.38} & {PDFTranslator} & {0.44} & {0.48} & {Synapse} & {0.33} & {0.33} & {WorkFlow} & {0.27} & {0.27} \tabularnewline
 & {arc} & {0.15} & {0.15} & {poi-3.0} & {0.38} & {0.37} & {SystemData} & {0.25} & {0.24} & {Xalan} & {0.40} & {0.10} \tabularnewline
{LOC} & {Ivy} & {0.29} & {0.29} & {Prop} & {0.16} & {0.16} & {TermoProjekt} & {0.48} & {0.48} & {Xerces} & {0.43} & {0.32} \tabularnewline
 & {jEdit} & {\cellcolor{gray!40}0.05} & {\cellcolor{gray!40}0.05} & {Redaktor} & {\cellcolor{gray!40}0.02} & {\cellcolor{gray!40}0.02} & {Tomcat} & {0.26} & {0.26} &  &  &  \tabularnewline
 & {log4j} & {0.18} & {\cellcolor{gray!40}0.01} & {Serapion} & {0.35} & {0.35} & {Velocity} & {0.24} & {0.23} &  &  &  \tabularnewline
\hline 
\end{tabular}{\scriptsize\par}
\label{tab:correlation_ken_JURE}}
\end{table}
\end{landscape}

The results show that class size is significantly correlated with several of our OO metrics.
LOC is also significantly correlated with the number of defects. It was also found that all of the OO metrics except DIT are strongly correlated with the number of defects and defect-proneness \footnote{all associations are statistically significant at an $p < 0.05$.}.
We then used regression to fit models that would enable us to assess the mediation and moderation effects of size on the number of defects. In this analysis regression is used to explain the strength of the effect of independent variables on the dependent variable, rather than to predict outcomes. We used each OO metric as an independent variable and the number of defects as the dependent variable. 



\subsection{The Mediation Effect of Class Size (RQ1)}
\label{sec:mediationAnlaysis}
Having confirmed that the OO metrics, size metric and defects are all significantly correlated, and that some of these OO metrics can explain (to varying degrees) the number of defects, we turn our attention to investigate whether size plays a mediating role in the relationships between the OO metrics and the number of defects.

\subsubsection{Mediation Effect On the Number Of Defects(RQ1.1)}
\label{sec:mediationAnlaysis:defects}

As we explained in Section \ref{sec:settings:medanalysis}, the mediation effect of variables is calculated based on three sequential steps - first by calculating the \textit{total effect} $c$ (Fig. \ref{fig:pathDiagram_normal}), then the \textit{direct effect} $c'$, and finally the \textit{indirect effect} $ab$ (Fig. \ref{fig:pathDiagram_mediation}). 
Note that we did not conduct any mediation or moderation analysis for DIT. While DIT is significantly correlated with the number of defects (with a rather weak negative correlation), it does not significantly explain the number of defects. 
This violates the first and second steps of our three-step procedure. Therefore, DIT was eliminated from any further analysis.

As there are results from 23 different systems acress two different datasets, we present here a detailed explanation from only two selected systems (one taken from each dataset), and we include detailed results for all other individual systems in our online repository \citep{repository}

We present the results of the mediation analysis of class size on the number of defects from two systems: Eclipse Mylyn (DAMB) and Apache Xerces (JURE). These two systems were selected because they are well-known, developed by well-established communities and they are two of the largest systems in each of the datasets.  
The results of the mediation effect of class size on the relationship between OO metrics and the number of defects in Mylyn are presented in Table \ref{tab:med_count_mylyn}. The \textit{total effect} ($c$) regression model for RFC explaining the number of defects, ignoring the mediator, was significant ($p<0.001$, Estimate= 1.68e-03). The direct effect ($c'$) of RFC on defects - by introducing the mediator LOC as a further explanatory factor in the model - was insignificant ($p<0.91$, Estimate= -3.49e-05). Failing in this second step suggests that a mediation effect of class size is unlikely, although the indirect effect was significant ($p<0.001$, Estimate= 1.71e-03), but quite weak. 
The same inference applies to the WMC, LCOM and Fan-in metrics (i.e., the total and indirect effects are significant, but the direct effect is insignificant). Therefore, it is concluded that class size \emph{does not fully mediate} the relationship between the number of defects and these metrics: RFC, WMC, LCOM, Fan-in and Fan-out in Mylyn. 
On the other hand, for CBO, the total effect ($c$) of CBO on the number of defects was found to be significant ($p<.001$, Estimate= 0.0027). The direct effect ($c'$), as we introduce the mediator LOC in the model, was also significant ($p<.001$, Estimate= 0.0013). We then estimate the indirect effect which was also significant ($p<.001$, Estimate= 0.0014) - the indirect effect of CBO on defects interval does not include zero (LLCI= 0.0009, ULCI= 0.0001); i.e., zero does not fall between the bootstrapping LLCI and ULCI values. The same applies to the Fan-out metric, which exhibits a significant total, direct and indirect effects (all at $p<0.001$). Therefore, it is inferred that \emph{size fully mediates} the relationship between CBO and Fan-out, and the number of defects.

\begin{table}[h]
\caption{The mediation effect of class size with the number of defects in Mylyn}
\label{tab:med_count_mylyn}
\begin{tabular}{@{}lccccccc@{}}
\toprule
\multirow{2}{*}{Metrics} & \multicolumn{2}{c}{\textbf{Indirect Effect}}   & \multicolumn{2}{c}{\textbf{Direct Effect}}     & \multicolumn{2}{c}{\textbf{Total Effect}}      & \multicolumn{1}{c}{\multirow{2}{*}{Significant}} \\  \cmidrule(lr){2-7}
                          & Estimate & p-value                   & Estimate  & p-value                   & Estimate & p-value                   & \multicolumn{1}{c}{}                             \\ \midrule
RFC                      & 1.71e-03  & \textbf{\textless{}0.001} & -3.49e-05  & 0.91 & 1.68e-03  & \textbf{\textless{}0.001} &                                                 \\
WMC                      & 0.004267 & \textbf{\textless{}0.001}                     & -0.000529  & 0.29                     & 0.003737 & \textbf{\textless{}0.001} &                                                  \\
CBO                      & 0.005006 & \textbf{\textless{}0.001} & 0.001439  & \textbf{\textless{}0.002} & 0.006445 & \textbf{\textless{}0.001} &   \Checkmark                                               \\
LCOM                     & 3.43e-04  & \textbf{\textless{}0.001} & -3.87e-05 & 0.37                      & 3.04e-04 & \textbf{\textless{}0.001}  &                                                  \\
Fan-in                   & 0.003800 & \textbf{\textless{}0.01}           & 0.001145  & 0.10  & 0.004945 & \textbf{\textless{}0.001} &                                                \\ 
Fan-out                  & 0.00854  & \textbf{\textless{}0.001}  & 0.00716  & \textbf{\textless{}0.001} & 0.01570 & \textbf{\textless{}0.001} & \Checkmark             \\ \bottomrule
\end{tabular}
\end{table}

The results of the mediation analysis for Xerces (JURE dataset) are shown in Table \ref{tab:med_count_xerces}. The regression model that uses RFC to explain the variance in the number of defects while ignoring the mediator ( \textit{total effect}) was significant ($p<0.001$, Estimate= 0.0325). The \textit{direct effect} ($c'$) of RFC on defects - by introducing the mediator ($\log(LOC)$) as a further explanatory factor in the model - is also significant ($p<0.001$, Estimate= 0.0113). A concurring outcome was evident for the \textit{indirect effect} of RFC on the number of defects, which was also
significant ($p<0.001$, Estimate= 0.0212) and zero does not fall between the interval values (LLCI= .0.013, ULCI= 0.03), which indicates that (unlike the results in Mylyn) size actually \textit{mediates} the relationship between RFC and number of defects. This is the same for the following metrics: CBO (total effect= 0.1008, direct effect= 0.0279, indirect effect = 0.0630, all significant at $p<0.001$), Fan-in (total effect= 0.0993, direct effect= 0.02959, indirect effect = 0.0697, $p<0.05$) and Fan-out (total effect= 0.1783, direct effect= 0.0671, indirect effect = 0.1112, significant at $p<0.001$). These results indicate that class size indeed \textit{mediates} the relationships between RFC, CBO, Fan-in and Fan-out and the number of defects in Xerces.
However, for the other two metrics for Xerces (i.e., WMC and LCOM), class size does \textit{not} mediate their relationships with the number of defects. For example, WMC shows a significant total effect ($p<0.001$, Estimate= 0.0744) but an insignificant direct effect ($p<0.21$, Estimate= 0.0099), suggesting that the mediation effect of size is unlikely.

\begin{table}[]
\caption{The mediation effect of class size with the number of defects in Xerces}
\label{tab:med_count_xerces}
\begin{tabular}{@{}lccccccc@{}}
\toprule
\multirow{2}{*}{Metrics} & \multicolumn{2}{c}{\textbf{Indirect Effect}}   & \multicolumn{2}{c}{\textbf{Direct Effect}}     & \multicolumn{2}{c}{\textbf{Total Effect}}      & \multicolumn{1}{c}{\multirow{2}{*}{Significant}} \\  \cmidrule(lr){2-7}
                         & Estimate  & p-value                   & Estimate  & p-value                   & Estimate  & p-value                   & \multicolumn{1}{c}{}                             \\  \midrule
RFC                      & 0.02124  & \textbf{\textless{}0.001} & 0.01129  & \textbf{\textless{}0.001} & 0.03253  & \textbf{\textless{}0.001} &     \Checkmark                                             \\
WMC                      & 0.06455  & \textbf{\textless{}0.001} & 0.00988  & 0.21                      & 0.07442  & \textbf{\textless{}0.001} &                                                  \\
CBO                      & 0.0630   & \textbf{\textless{}0.001} & 0.0379   & \textbf{\textless{}0.001} & 0.1008   & \textbf{\textless{}0.001} &   \Checkmark                                               \\
LCOM                     & 0.003550 & \textbf{\textless{}0.001} & 0.000439 & 0.59                      & 0.003989 & \textbf{\textless{}0.001} &                                                  \\
Fan-in                   & 0.06969  & \textbf{\textless{}0.001} & 0.02959  & \textbf{\textless{}0.05}  & 0.09928  & \textbf{\textless{}0.001} &  \Checkmark                                                \\
Fan-out                  & 0.1112   & \textbf{\textless{}0.001} & 0.0671   & \textbf{\textless{}0.001} & 0.1783   & \textbf{\textless{}0.001} &  \Checkmark                                                \\ \bottomrule
\end{tabular}
\end{table}

\begin{table}[h]
\centering
\caption{Summary of the mediation effect of class size in count models}
  \begin{adjustbox}{max width=\textwidth}

\begin{threeparttable}
\begin{tabular}{@{}lllccccc@{}}
\toprule
\multicolumn{2}{l}{Systems}                & RFC   & WMC   & CBO   & LCOM & Fan-in & Fan-out \\ \midrule
\multirow{5}{*}{\rotatebox[origin=c]{90}{\textbf{DAMB}}}\  & JDT             &  \Checkmark   &  \Checkmark   &  \Checkmark   & \Checkmark    &     & \Checkmark      \\ 
                       & Mylyn           &    &     &  \Checkmark   &    &    &   \Checkmark    \\
                       & PDE             &     &     &   &   &      &     \\
                       & Equinox         &     & \Checkmark    & \Checkmark   &  \Checkmark   & \Checkmark     &    \\
                       & Lucene          &    &     &  \Checkmark     &    & \Checkmark        &   \Checkmark      \\ \midrule
\multirow{18}{*}{\rotatebox[origin=c]{90}{\textbf{JURE}}} & Ant             & \Checkmark  &    &     &    &      & \Checkmark      \\
                       & Ivy             &  \Checkmark   &     &   \Checkmark  &    &  \Checkmark  &  \Checkmark     \\
                       & jEdit           &     &     &     &    &      &       \\
                       & Log4j           &  &     &   \Checkmark  &    &   &       \\
                       & POI             &  \Checkmark   &  & \Checkmark   &    &      & \Checkmark      \\
                       & Prop            &     &     &     &  \Checkmark  &      &       \\
                       & Synapse         &  \Checkmark   &     &  \Checkmark  &    &      &  \Checkmark    \\
                       &  Tomcat         &   \Checkmark  &     &  \Checkmark &   &  & NA      \\
                       &   Velocity      &   &   &  \Checkmark &   &    & \Checkmark   \\
                       & Xalan           &  \Checkmark  &    & \Checkmark &    &      &  \Checkmark    \\
                       & Xerces          &   \Checkmark  &    & \Checkmark   &    &   \Checkmark  &  \Checkmark    \\
                       & Arc             &    &    &    &   &     &       \\
                       & PDFTranslator   &     &     &     &    &     &      \\
                       & Redaktor        &    &   &   &   &     &       \\
                       & Serapion        &    &    & \Checkmark    &   &      &       \\
                       & SystemDataMangt &     &     &     &    &     &       \\
                       & TermoProjekt    &     &     &     &    &      &      \\
                       & WorkFlow        &     &     &  &    &      &      \\ \bottomrule
\end{tabular}
 \begin{tablenotes}
    \small
          \item \Checkmark the \textit{mediation} effect of size is significant.

          \item NA: metric data is not available. 
        \end{tablenotes}
     \end{threeparttable}
      \end{adjustbox}
      \label{tab:mediation_individualSystems_summary_conunt}
\end{table}

The same analysis process was enacted on all 23 systems. For enhanced readability, we provide only summary results of this analysis in Table \ref{tab:mediation_individualSystems_summary_conunt} (though detailed results from each system are available externally \citep{repository}). The table shows a summary of whether class size has a mediation effect for each single metric. As shown in Table \ref{tab:mediation_individualSystems_summary_conunt}, there are mixed results regarding the true mediation effect of class size across all systems examined. Considering the results for the DAMB dataset, we observed that size appears to have a mediation effect on the relationship between CBO and the number of defects in 4/5 systems (all systems except PDE), and on Fan-out in 3/5 systems. For the other metrics, there is no consistent evidence of a mediation effect for class size, given that there was no significant mediation effect in more than two systems. 

A similar pattern was found in the analysis of the JURE dataset, where only a few systems show a significant effect of class size on the relationship between metrics and the number of defects. Some systems such as JDT, Ivy and Xerces show a mediation effect of size for 4 or 5 metrics, where other systems such as Arc and PDFTranslator do not show a mediation effect for any of the metrics examined.
The mediation effect of size on the relationship between CBO and defects appears in 9/18 systems. A number of systems also show a mediation effect of size on a other coupling metrics, i.e., Fan-out and RFC, where the mediation effect of size appears in 7 systems each.
Another interesting observation from the results is that, for all the proprietary systems we included from the JURE dataset, none of the metrics seemed to be mediated by size (except for CBO in Serapion). However, given the still limited number of systems we investigated and the size of these systems we are not able to generalise these findings to other proprietary systems. Further investigation is needed for such systems. 


Overall, WMC, LCOM and Fan-in are the metrics with the least evidence of a mediation effect of size. We could not find evidence of a significant mediation effect of size on the relationship between those two metrics and the number of defects in the majority of systems we investigated (only in 2, 3 and 4 systems, respectively). The mediation effect on all six metrics seems to be different between systems as well.
When comparing the results we obtained by analyzing the DAMB dataset with those found prior in the study of 
\cite{Zhou2014} (using the same dataset), we found similar evidence of a mediation effect of size for CBO only, as the evidence of the mediation effect of size was consistent in 4 of the 5 systems. However, unlike the previous analysis \citep{Zhou2014}, we could not confirm if the effect of size is significant for the other metrics: RFC, WMC, LCOM, Fan-in or Fan-out metrics, simply because the results were largely inconsistent across different systems.

In summary, it is observed that evidence regarding the mediation effect of size is \textit{inconsistent} and does not follow a pattern across all systems. For all metrics, class size is not shown to have a significant mediation effect across all systems. Only CBO shows a more consistent significant mediation effect of size than other metrics, where a significant indirect effect was found in 13 out of 23 systems. There is less evidence of the effect of size on Fan-out, where the mediation effect is significant in 10 systems across both datasets. However, we still consider these results inconclusive in determining if the indirect effect is truly and always significant.
In returning to our hypothesis ({H01.1}), based on the evidence resulting from our analysis of the individual systems in both datasets, we reject the null hypothesis that class size has no mediation effect on the relationship between the number of defects and the following OO metrics: RFC, WMC, LCOM, Fan-in metrics. For CBO, while the mediation effect appears in the majority of systems, the evidence of this effect is more consistent in DAMB dataset (4/5) than in JURE dataset (9/18).

\subsubsection{Mediation Effect on Defect-Proneness (RQ1.2)}
\label{sec:mediationAnlaysis:defect-pronenss}
Having investigated the effect of class size on the number of defects in the previous Section (\ref{sec:mediationAnlaysis:defects}), we now investigate the mediation effect of class size on the presence of defects in software modules (e.g., defect-proneness models). Similar to the previous section, we present the results of mediation analysis for a system from each dataset: Eclipse Mylyn and Apache Xerces in Tables \ref{tab:med_binary_mylyn} and \ref{tab:med_binary_xerces}, respectively.

\begin{table}[h]
\centering

\caption{The mediation effect of class size with defect-proneness in Mylyn}
\label{tab:med_binary_mylyn}
\begin{tabular}{@{}lccccccc@{}}
\toprule
\multirow{2}{*}{Metrics} & \multicolumn{2}{c}{\textbf{Indirect Effect}}   & \multicolumn{2}{c}{\textbf{Direct Effect}}     & \multicolumn{2}{c}{\textbf{Total Effect}}      & \multicolumn{1}{c}{\multirow{2}{*}{Significant}} \\  \cmidrule(lr){2-7}
                         & Estimate  & p-value                   & Estimate  & p-value                   & Estimate  & p-value                   & \multicolumn{1}{c}{}                             \\  \midrule
RFC                      & 4.51e-04 & \textbf{\textless{}0.001} & 2.20e-04  & 0.11            & 6.71e-04 & \textbf{\textless{}0.001} &                                                  \\
WMC                      & 1.33e-03 & \textbf{\textless{}0.001} & -2.92e-05 & 0.97                      & 1.31e-03 & \textbf{\textless{}0.001} &                                                  \\
CBO                      & 0.001426 & \textbf{\textless{}0.001} & 0.001298  & \textbf{\textless{}0.001} & 0.002724 & \textbf{\textless{}0.001} & \checkmark                                        \\
LCOM                     & 1.06e-04 & \textbf{\textless{}0.001} & -2.40e-05 & 0.31                      & 8.19e-05 & \textbf{\textless{}0.01}  &                                                  \\
Fan-in                   & 0.001076 & \textbf{\textless{}0.001} & 0.000744  & 0.09                      & 0.001821 & \textbf{\textless{}0.01}  &                                                  \\
Fan-out                  & 0.001349 & 0.11                      & 0.006553  & \textbf{\textless{}0.001} & 0.007903 & \textbf{\textless{}0.001} &                                                  \\\bottomrule
\end{tabular}
\end{table}

For the mediation analysis conducted on Mylyn (Table \ref{tab:med_binary_mylyn}), the \textit{total} and \textit{indirect effects} of RFC on defect-proneness intervals were significant and zero does not lie between the LLCI and ULCI (indirect effect: $p<0.01$, Estimate= 4.51e-04), however, the \textit{direct effect} ($c'$) regression model for RFC explaining defect-proneness was not significant ($p<0.11$, Estimate= 4.51e-04), which inferred that size does not fully mediate the relationship between RFC and defect-proneness. The same conclusion is observed for the relationships between defect-proneness and WMC (\textit{direct effect}: $p<0.97$, Estimate= -2.92e-05), LCOM (\textit{direct effect}: $p<0.31$, Estimate= -2.40e-05) and Fan-in (\textit{direct effect}: $p<0.09$, Estimate= 0.0007=). For Fan-out, while both the \textit{direct} and \textit{total effects} were found to be significant at $p<0.001$ (with an Estimate of 0.0066 and 0.0079, respectively), the \textit{indirect effect} was insignificant ($p<0.11$, Estimate= 0.0013, LLCI = -0.0003, ULCI = 0). This also indicates the the mediation effect of size is questionable.
On other hand, we observed a mediation effect of size on the relationship between CBO and defect-proneness in Mylyn as all three effects were found to be significant at $p<0.001$. The indirect effect was significant with Estimate= 0.0014. This observation shows that size has a significant mediation effect on the relationship between CBO and class defect-proneness.

\begin{table}[h]
\caption{The mediation effect of class size with defect-proneness in Xerces}
\label{tab:med_binary_xerces}
\begin{tabular}{@{}lccccccc@{}}
\toprule
\multirow{2}{*}{Metrics} & \multicolumn{2}{c}{\textbf{Indirect Effect}}   & \multicolumn{2}{c}{\textbf{Direct Effect}}     & \multicolumn{2}{c}{\textbf{Total Effect}}      & \multicolumn{1}{c}{\multirow{2}{*}{Significant}} \\  \cmidrule(lr){2-7}
                         & Estimate  & p-value                   & Estimate  & p-value                   & Estimate  & p-value                   & \multicolumn{1}{c}{}                             \\  \midrule
RFC                      & 0.02124  & \textbf{\textless{}0.001} & 0.01129  & \textbf{\textless{}0.001} & 0.03253  & \textbf{\textless{}0.001} & \checkmark                                        \\
WMC                      & 0.06455  & \textbf{\textless{}0.001} & 0.00988  & 0.21                      & 0.07442  & \textbf{\textless{}0.001} &                                                  \\
CBO                      & 0.0630   & \textbf{\textless{}0.001} & 0.0379   & \textbf{\textless{}0.001} & 0.1008   & \textbf{\textless{}0.001} & \checkmark                                        \\
LCOM                     & 0.003550 & \textbf{\textless{}0.001} & 0.000439 & 0.59                      & 0.003989 & \textbf{\textless{}0.001} &                                                  \\
Fan-in                   & 0.06969  & \textbf{\textless{}0.001} & 0.02959  & \textbf{\textless{}0.05}  & 0.09928  & \textbf{\textless{}0.001} & \checkmark                                        \\
Fan-out                  & 0.1112   & \textbf{\textless{}0.001} & 0.0671   & \textbf{\textless{}0.001} & 0.1783   & \textbf{\textless{}0.001} & \checkmark                                        \\  \bottomrule
\end{tabular}
\end{table}

In Xerces (Table \ref{tab:med_binary_xerces}), for the RFC, CBO,  Fan-in and Fan-out metrics, all three \textit{direct}, \textit{indirect} and \textit{total effects} regression models for those metrics explaining defect-proneness were significant at $p<0.001$. For example, size has an \textit{indirect effect} on the relationship between CBO and defect-proneness with $p<0.001$ and Estimate= 0.0630. The mediation effect of class size is considered to be significant for four metrics. However, for WMC and LCOM, while the \textit{total} and  \textit{indirect} effects were significant, the  \textit{direct effect} of size was insignificant (violating the second step). In the case of LCOM, 
the \textit{indirect effect} is insignificant with $p<0.59$, and Estimate= 0.0004, indicating that the mediation effect of class size is improbable.

\begin{table}[h]
\centering
\caption{Summary of the mediation effect of class size in binary models}
  \begin{adjustbox}{max width=\textwidth}

\begin{threeparttable}
\begin{tabular}{@{}lllccccc@{}}
\toprule
\multicolumn{2}{l}{Systems}                & RFC   & WMC   & CBO   & LCOM & Fan-in & Fan-out \\ \midrule
\multirow{5}{*}{\rotatebox[origin=c]{90}{\textbf{DAMB}}} & JDT             &  \Checkmark   &   \Checkmark    &  \Checkmark   &   \Checkmark  &       &   \Checkmark    \\
     & Mylyn           &    &      &  \Checkmark   &     &      &        \\
      & PDE             &    &    &      &    &       &       \\
     & Equinox         &   &   \Checkmark   & \Checkmark  &  \Checkmark   &   \Checkmark   &        \\
     & Lucene          &   &      &    \Checkmark  &     &    \Checkmark   & \Checkmark       \\ \midrule
\multirow{18}{*}{\rotatebox[origin=c]{90}{\textbf{JURE}}} & Ant             &  \Checkmark &      &      &    &       &   \Checkmark    \\
     & Ivy             &   &      &      &     &       &        \\
     & jEdit           &    &      &      &     &       &        \\
     & Log4j           &    &      &      &     &       &        \\
     & POI             &    &      & \Checkmark   &     &       &       \\
     & Prop            &    & \Checkmark   & \Checkmark     & \Checkmark     &       &        \\
     & Synapse         &    &      & \Checkmark    &     &       &        \\
     & Tomcat          &  \Checkmark  &      &  \Checkmark   &     &    & NA       \\
     & Velocity        &    &  \Checkmark     &  \Checkmark     & \Checkmark     &       &       \\
     & Xalan           &  \Checkmark & \Checkmark   &   \Checkmark   &   \Checkmark   &       & \Checkmark       \\
     & Xerces          & \Checkmark   &    & \Checkmark     &     &   \Checkmark    &   \Checkmark     \\
     & Arc             &  &     &     &    &       &        \\
     & PDFTranslator   &    &      &   \Checkmark &  \Checkmark   &       &     \\
     & Redaktor        &   &      &   &    &   &        \\
     & Serapion        &  &  & \Checkmark & \Checkmark    &  &       \\
     & SystemDataMangt &   &     &     &    &      &       \\
     & TermoProjekt    &  &  &      &     &      &    \\
     & WorkFlow        & &  &    &     &       &        \\ \bottomrule
\end{tabular}
\begin{tablenotes}
    \small
             \item \Checkmark  the \textit{mediation} effect of size is significant.

          \item NA: metric data is not available. 
        \end{tablenotes}
     \end{threeparttable}
      \end{adjustbox}
      \label{tab:mediation_individualSystems_summary_binary}
\end{table}

When looking at the two datasets combined (Table \ref{tab:mediation_individualSystems_summary_binary}), we observed that the mediation effect of size is significant for CBO in 13 systems, with particularly strong evidence in the DAMB dataset (being evident in all systems except PDE). The size effect on CBO is also observed in 9 other systems in JURE dataset. This is similar to the pattern reported for the count models in Section \ref{sec:mediationAnlaysis:defects}. 
None of the other metrics have shown similar/consistent results in either of the datasets. Unlike the results from the count models, there is less evidence of the mediation effect of size on RFC and Fan-out (i.e., significant in only 5 systems across both datasets). LCOM is the metric with the second highest number of systems where the effect of size is found to be significant (found in 7/23 systems). Therefore, the results from all metrics except CBO are considered to be diverse and do not follow a consistent pattern across all datasets/systems.

To summarise the results from the defect-proneness analysis, which are consistent with the results from the count defect models, it is observed that evidence regarding the mediation effect of class size is still discordant and does not follow a specific pattern across all systems. Again, only CBO shows a more consistent result, as size appears to have indirect effects on the relationship between CBO and class defect-proneness in 13 systems (4 systems in DAMB and 9 systems in JURE). Therefore, we consider the mediation effect of class size on defect-proneness to have a potential impact only on the relationship between CBO and defect-proneness. In returning to our research hypothesis ({H01.2}), we reject the null hypothesis for the following metrics RFC, WMC, LCOM, Fan-in and Fan-out but we cannot reject the null hypothesis for CBO as the results are mixed and we could not find strong evidence of the absence of the mediation effect of class size.

When looking at the results from both sets of models- i.e., count (Section \ref{sec:mediationAnlaysis:defects}) and binary (Section \ref{sec:mediationAnlaysis:defect-pronenss}), we observe that only CBO resulted in the same significant mediation effect of class size on the relationship with defects/defect-proneness. Eleven (out of thirteen) of the systems that show a mediation effect of size in the count models also show a mediation effect in the binary models. This suggests that CBO is indeed the metric that is mostly impacted by the mediation effect of class size. Our results demonstrate that there is either a lack of statistical evidence (or in many cases, no strong evidence) of a significant mediation effect of size in defect count and binary defect prediction models. The only metric that looks vulnerable to the effect of size is CBO.

\subsection{The Moderation Effect of Class Size (RQ2)}
\label{sec:moderationAnlaysiss}
We now look at the moderation (interaction) effect of class size. Similar to the mediation analysis, we study the moderation effect of size in models explaining both the number of defects (count) and defect-proneness (binary). 
To investigate the moderation effect of class size on the number of defects, we empirically assessed the impact of all six OO metrics on the number of defects under the influence of class size, and report the results for two systems separately (one selected from each dataset to enhance readability) - Mylyn and Xerces. 
As explained in Section \ref{sec:medmodmethods}, the moderation effect is estimated using multiple regression analysis. 
We considered the moderation effect of class size on the relationship between individual OO metrics and the number of defects/defect-proneness to be significant when the interaction effect is significant ($p<.05$).

To graphically illustrate the moderation effect of class size on the relationship between the object-oriented metrics and defects,   three levels of the moderator (mean, one standard deviation below the mean (-SD) and one standard deviation above the mean (+SD)) are computed and then plotted using simple slopes. The plots are generated using the rockchalk function \citep{johnson2012rockchalk} available in \texttt{R}.


\subsubsection{Moderation Effect on the Number of Defects (RQ2.1)}
\label{sec:mod_cont}
Table \ref{tab:tabmod} presents the moderation effect results for Mylyn and Xerces from the DAMB and JURE datasets, respectively.
In Table \ref{tab:tabmod}, our moderation analysis reveals that the interaction effects  of size on defects for Mylyn were significant for four metrics (RFC, LCOM, Fan-in and Fan-out), all with significant interaction terms ($p<0.05$). For WMC and CBO, this indicates that there is no significant relationship between these metrics and the number of defects. 
Similarly, for Xerces, we observed significant moderation effects of LOC on the individual OO metrics excluding the LCOM ($p = 0.229$) and Fan-in ($p = 0.452$) metrics. Looking at both systems, the moderation effect of size appears to be evident only for the RFC and Fan-out metrics.

\begin{table}[htb]
\centering
\caption{Moderation effect of size on the relationship between individual OO metrics and defects for Mylyn (DAMB Dataset) and Xerces (JURE Dataset): count models}
\begin{tabular}{lccc|ccc}
\hline 
\multicolumn{4}{c|}{Eclipse Mylyn} & \multicolumn{3}{c}{Apache Xerces}\tabularnewline
\hline 
Metrics & Estimate & p-value & Significant & Estimate & p-value & Significant\tabularnewline
\hline 
RFC & -0.0023606 & \textbf{\textless{}0.001} & \checkmark  & -0.0025988 & \textbf{\textless{}0.001} & \checkmark  \tabularnewline
WMC & -0.0005414 & 0.425 &  & 0.004101 & \textbf{\textless{}0.001} & \checkmark \tabularnewline
CBO & 0.002356 & 0.122  &  & 0.003280 & \textbf{\textless{}0.01} & \checkmark \tabularnewline
LCOM & -1.695e-04 & \textbf{\textless{}0.01} & \checkmark & -7.379e-05 & 0.229 & \tabularnewline
Fan-in & 0.005272 & 0.0481 & \checkmark & 0.001427 & 0.452 & \tabularnewline
Fan-out & -0.009447 & \textbf{\textless{}0.01} & \checkmark  & -0.009595 & \textbf{\textless{}0.001} & \checkmark  \tabularnewline
\hline 
\end{tabular}
\label{tab:tabmod}
\end{table}

\begin{figure}[h]
\centering
\includegraphics[width=0.60\linewidth]    {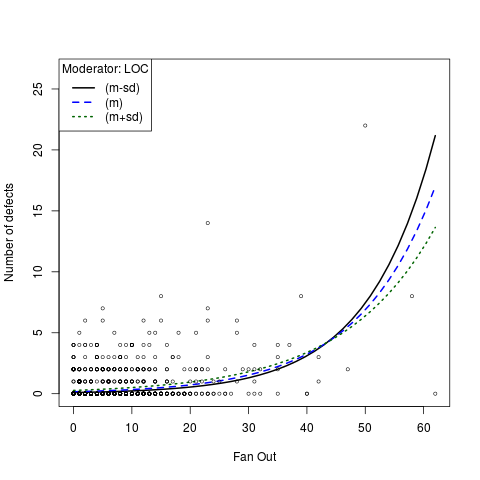}
    \caption{Moderation effect of size on the relationship between Fan-out and the number of defects in Eclipse Mylyn}
     \label{fig:Moderation_fanout_DS1}
\end{figure}

\begin{figure}[h]
\centering
\includegraphics[width=0.60\linewidth]
    {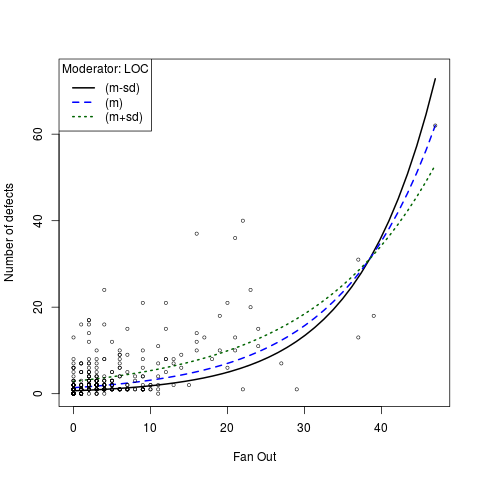}
    \caption{Moderation effect of size on the relationship between Fan-out and the number of defects in Apache Xerces}
     \label{fig:Moderation_fanout_DS2}
\end{figure}

We then present examples of the moderation effects of size on the relationship between Fan-out and defects for Eclipse Mylyn (DAMB dataset) in Fig. \ref{fig:Moderation_fanout_DS1}, and Fan-out and defects for Apache Xerces (JURE dataset) in Fig. \ref{fig:Moderation_fanout_DS2}. We include  graphs from all other systems in our online repository \citep{repository}. 

As shown in Fig. \ref{fig:Moderation_fanout_DS1}, we observe that small class size values (black line) with small Fan-out values resulted in  identifying more defects 
. 
Large size values (green dotted line) together with small Fan-out values initially resulted in identifying more defects than the average (blue dashed line). However, as the Fan-out values increased, the number of defects identified by the small and large size values interchanged - i.e., more defects were identified than average when size was small, and more defects were identified also when size was large, but it was less than the number identified by the average size values.
Regarding the Xerces results, the simple slope plot in Fig. \ref{fig:Moderation_fanout_DS2} shows a similar trend to that of the Mylyn project in Fig. \ref{fig:Moderation_fanout_DS1}. From Fig. \ref{fig:Moderation_fanout_DS2}, we observe that small size values (the black line) leads to more defects being found with large Fan-out values but leads to less defects found overall on average (the blue dashed line). Large size values (the green dotted line) with large Fan-out values help find more defects than average. However, as Fan-out increases beyond 35, the number of defects identified by the small size and large size values interchange where the small size values identify more defects than the large size values. This suggests that the Fan-out metric has a high probability of explaining defects count in large classes (where LOC is high). The difference in the slopes for small and large values of size confirms that size does moderate the relationship between Fan-out and the number of defects.

The moderation analysis results for all systems in all datasets are shown in Table \ref{tab:mod_individualSystems_summary_count}. We only provide the summarised results, which show whether size has a moderation effect on the relationship between each single metric and defects. Overall, the results were mixed. Significant moderation effects are observed for all metrics for Lucene. Three systems show significant moderation effects for four metrics (i.e., Mylyn, jEdit and Xerces), although for different combinations of metrics. On the other hand, seven systems (i.e., Equinox, Ivy, Log4j, Velocity, PDFTranslator, TemoProjekt and Workflow) show no moderation effect of class size on any of the examined metrics (TemoProjekt and Workflow also shows no mediation effect - see Tables \ref{tab:mediation_individualSystems_summary_conunt} and \ref{tab:mediation_individualSystems_summary_binary}). The moderation effect of size on RFC and Fan-out metrics was significant in 10 systems across both datasets, while the moderation effect of size on LCOM was significant in only five systems. The moderation effect of size on the other four remaining metrics was significant in fewer than 5 systems. Class size seems to moderate the relationship between LCOM and defects in 3/5 systems in the DAMB dataset, but this was not the case in the JURE dataset (only 2/18 systems). For Fan-in, the moderation effect was also significant in two systems in the DAMB dataset but was significant in only two of the eighteen systems in the JURE dataset. 
While we have found evidence of a \textit{mediation} effect of size on the relationship between CBO and the number of defects, we can confirm that there is no \textit{moderation} effect of size for the same metric.  

\begin{table}[h]
\centering
\caption{Summary of the moderation effect of class size in count models}
  \begin{adjustbox}{max width=\textwidth}

\begin{threeparttable}
\begin{tabular}{@{}lllccccc@{}}
\toprule
\multicolumn{2}{l}{Systems}                & RFC   & WMC   & CBO   & LCOM & Fan-in & Fan-out \\ \midrule
\multirow{5}{*}{\rotatebox[origin=c]{90}{\textbf{DAMB}}} & JDT             & NA    &     &     &      &       & \Checkmark      \\
     & Mylyn           & \Checkmark    &     &    & \Checkmark   & \Checkmark     & \Checkmark      \\
      & PDE             & \Checkmark    &     &      & \Checkmark    &       &       \\
     & Equinox         &   &    &  &     &      &        \\
     & Lucene          & \Checkmark   & \Checkmark     & \Checkmark     & \Checkmark     & \Checkmark      & \Checkmark        \\ \midrule
\multirow{18}{*}{\rotatebox[origin=c]{90}{\textbf{JURE}}} & Ant             & \Checkmark  &      &      &     &       &      \\
     & Ivy             &    &      &     &    &       &        \\
     & jEdit           & \Checkmark    & \Checkmark      &      & \Checkmark     &       & \Checkmark        \\
     & Log4j           &    &      &      &     &       &        \\
     & POI             & \Checkmark    &      &   & \Checkmark     &       & \Checkmark      \\
     & Prop            &   & \Checkmark  &      &    &       &        \\
     & Synapse         &    &      &     &    &       & \Checkmark      \\
     & Tomcat          & \Checkmark   &      & \Checkmark    &     &    & NA       \\
     & Velocity        &    &      &      &     &       &        \\
     & Xalan           & \Checkmark  &     &      &    &      & \Checkmark      \\
     & Xerces          & \Checkmark   & \Checkmark  & \Checkmark    &     &      & \Checkmark      \\
     & Arc             & \Checkmark   &      &      &     &       & \Checkmark       \\
     & PDFTranslator   &    &      &     &    &       &    \\
     & Redaktor        &    &      &     &    &       & \Checkmark       \\
     & Serapion        &    &   &      &     & \Checkmark      &       \\
     & SystemDataMangt &    &      &      &    & \Checkmark      &        \\
     & TermoProjekt    &    &     &     &    &       &        \\
     & WorkFlow        &    &      &     &     &       &        \\ \bottomrule
\end{tabular}
 \begin{tablenotes}
    \small
          \item \checkmark the \textit{moderation} effect of size is significant.
           \item NA: metric data is not available. 
        \end{tablenotes}
     \end{threeparttable}
      \end{adjustbox}
      \label{tab:mod_individualSystems_summary_count}
\end{table}

In summary, it is observed that evidence regarding a moderation effect of size is \textit{inconsistent} and does not follow a pattern across all systems. 
Only RFC and Fan-out show a consistent and more significant moderation effect of size than other metrics, where a significant moderation effect was found in 10 systems.
Returning to our research hypothesis ({H02.1}), we reject the null hypothesis that class size has no significant moderation effect on the relationship between the number of defects and the following OO metrics: WMC, CBO, LCOM and Fan-in. For RFC and Fan-out, the evidence available for this effect is not consistent across both datasets. Again, and similar to the mediation effect, we are unable to confirm if a moderation effect will always be present.
\\

\subsubsection{Moderation Effect on Defect-Proneness (RQ2.2)}
For binary models (i.e., models of defect-proneness), we apply the same analysis conducted for count models (see Section \ref{sec:mod_cont}). 
A significant $p<.05$ with positive estimate value indicates a significant and positive association between the metric and defect-proneness (i.e., an increase in the OO metric corresponds to an increase in the defect-proneness of modules) whereas a negative effect value indicates that the relationship between the metric and defects is negative (i.e., when the value of the OO metric increases, the defect-proneness decreases). For an insignificant p-value, it indicates that there is no significant relationship between the OO metric and defect-proneness. 

Table \ref{tab:tabmodbin} presents the moderation effect results for Eclipse Mylyn and Apache Xerces from the DAMB and JURE datasets, respectively. 
The moderation analysis conducted for the two systems reveals that the interaction effects for Mylyn were insignificant for all but two metrics (RFC and CBO). 
However, for Xerces, we observed significant moderation effects of size on the individual OO metrics excluding the CBO ($p = 0.68$) and Fan-in metrics ($p = 0.50$). 

\begin{table}[h]
\centering
\caption{Moderation effect of size on the relationship between individual OO metrics and defects for Mylyn (DAMB Dataset) and Xerces (JURE Dataset): binary models}
 \vspace{0.2cm}
 \begin{tabular}{lccc|ccc}
\hline 
\multicolumn{4}{c|}{Eclipse Mylyn} & \multicolumn{3}{c}{Apache Xerces}\tabularnewline
\hline 
Metrics & Estimate & p-value & Significant & Estimate & p-value & Significant\tabularnewline
\hline 
RFC & -0.004251 & \textbf{\textless{}0.01} & \checkmark  & 0.025058 & \textbf{\textless{}0.001} & \checkmark  \tabularnewline
WMC & -0.0004594 & 0.761 &  & 0.033668 & \textbf{\textless{}0.001} & \checkmark  \tabularnewline 
CBO & 0.007743 & 0.0331 & \checkmark & -0.01608 & 0.68074  & \tabularnewline
LCOM & -5.576e-05 & 0.695 &  & 0.004701 & \textbf{\textless{}0.001} & \checkmark  \tabularnewline
Fan-in & 0.007078 & 0.169 &  & -0.01890 & 0.500953 & \tabularnewline
Fan-out & -0.007564 & 0.38306 &  & -0.29659 & \textbf{\textless{}0.001} & \checkmark  \tabularnewline
\hline 
\end{tabular}
\label{tab:tabmodbin}
  \vspace{0.2cm}
\end{table}

 \begin{figure}[h]
\centering
 \vspace{-0.2cm}
\includegraphics[width=0.60\linewidth]    {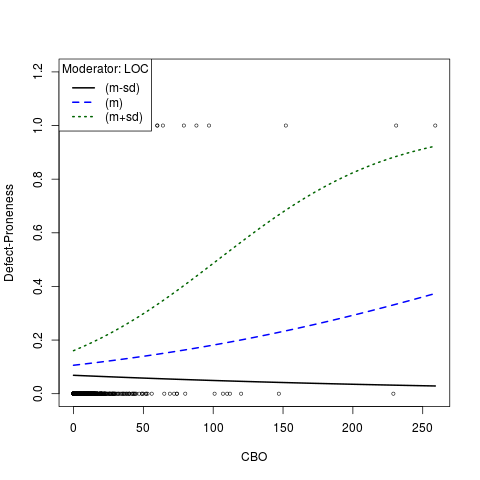}
    \caption{Moderation effect of size on the relationship between CBO and defect-proneness in Mylyn}
     \label{fig:Moderation_rfc_DS1}
\end{figure}

\begin{figure}[]
\centering
\includegraphics[width=0.60\linewidth]
    {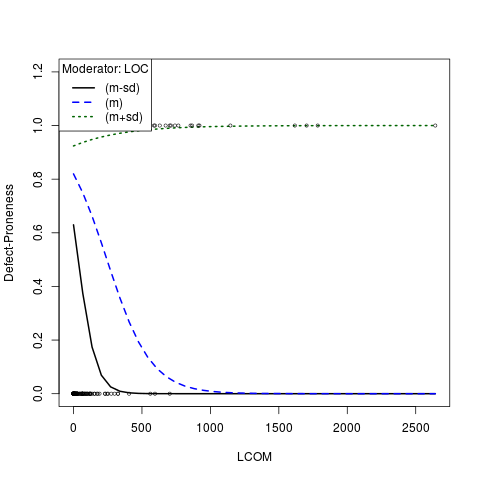}
    \caption{Moderation effect of size on the relationship between LCOM and defect-proneness in Xerces}
     \label{fig:Moderation_wmc_DS2}
\end{figure}

The simple slope analysis shown in Figs \ref{fig:Moderation_rfc_DS1}  and \ref{fig:Moderation_wmc_DS2} visually illustrates the moderation effects of size on the relationship between CBO and defect-proneness in Mylyn (DAMB dataset) and the moderation effect of size on the relationship between LCOM and defect-proneness in Xerces (JURE dataset), respectively. We provide all remaining graphs for all metrics from all other systems in our online repository \citep{repository}.

Fig. \ref{fig:Moderation_rfc_DS1} shows that small size values (the black line) with large CBO values leads to a less probability of defect-prone modules being found  in comparison to the average size values (the blue dashed line).  Large size values (the green dotted line) with large CBO values leads to a higher probability of defect-prone modules being found than average (the blue dashed line). Similarly, in Fig. \ref{fig:Moderation_wmc_DS2}, we observe that an initial increase in LCOM with small size values (black line) leads a less probability of defect-prone modules being found which is a less probability than average size  (the blue dashed line). However, large size values (green dotted line) with large LCOM values leads to higher probability of defect-prone modules being found which is more than the average. Furthermore, as the LCOM values increases, the probability of defect-prone modules  found for small size values and average size values are almost the same.  Also, as the LCOM values increase, the large size values leads to the same probability of defect-prone modules  being detected as previously detected by the small LCOM values. The difference in the slopes for large or small size values shows that size actually \textit{moderates} the relationship between LCOM and defect-proneness.


Table \ref{tab:mod_individualSystems_summary_binary} presents the summarised moderation analysis for all systems. Overall, the moderation effect of size on the relationship between the individual metrics and defect-proneness was insignificant for most of the systems. Only one system (JDT) shows significant moderation effects for 5/6 metrics. The moderation effect of size on Fan-out was significant in 9 systems across both datasets, while the effect on CBO was significant in only two systems. On the other hand, the effect on Fan-in was not significant for any of the systems. A further analysis of this effect across all systems revealed little significant moderation effect of size on the relationship between WMC, CBO and Fan-in metrics and defect-proneness. Class size seems to moderate the relationship between CBO and defect-proneness in two out of 5 systems in DAMB dataset, but this was not the case in the JURE dataset (0 out of the 18 systems). Indeed, the moderation effects were inconsistent across the individual systems, and 
consequently, we note the evidence regarding the moderation effect of size is not conclusive with regard to these datasets.
In summary, it is observed that evidence regarding the moderation effect of size is also \textit{inconsistent} across all systems. 
The moderation effect of size  was significant and considerably consistent on the relationship between only Fan-out and defect-proneness, although this is only found in 9 systems.

\begin{table}[tbh]
\centering
\caption{Summary of the moderation effect of class size in binary models}
  \begin{adjustbox}{max width=\textwidth}

\begin{threeparttable}
\begin{tabular}{@{}lllccccc@{}}
\toprule
\multicolumn{2}{l}{Systems}                & RFC   & WMC   & CBO   & LCOM & Fan-in & Fan-out \\ \midrule
\multirow{5}{*}{\rotatebox[origin=c]{90}{\textbf{DAMB}}} & JDT             & \Checkmark    & \Checkmark       & \Checkmark    & \Checkmark     &      & \Checkmark      \\
     & Mylyn           & \Checkmark    &      & \Checkmark    &    &     &        \\
      & PDE             & \Checkmark    &      &      &     &       & \Checkmark      \\
     & Equinox         &    &      &   &     &      &        \\
     & Lucene          &    &      &      & \Checkmark     &       & \Checkmark        \\ \midrule
\multirow{18}{*}{\rotatebox[origin=c]{90}{\textbf{JURE}}} & Ant             &   &     &      &     &      &      \\
     & Ivy             &    &      &      &     &       &        \\
     & jEdit           & \Checkmark    & \Checkmark      &      & \Checkmark     &       &        \\
     & Log4j           &    &      &      &     &       & \Checkmark        \\
     & POI             & \Checkmark    & \Checkmark      &   & \Checkmark     &       & \Checkmark      \\
     & Prop            &    &   &      &     &       &        \\
     & Synapse         &    &      &     &     &       & \Checkmark      \\
     & Tomcat          &   &      &     &     &    & NA       \\
     & Velocity        &    &      &      &     &       & \Checkmark       \\
     & Xalan           &   &     &      &     &       &       \\
     & Xerces          & \Checkmark   & \Checkmark  &     & \Checkmark    &      & \Checkmark      \\
     & Arc             &    &      &      &     &       &        \\
     & PDFTranslator   &    &      &     &     &       &     \\
     & Redaktor        &    &      &      &     &       & \Checkmark       \\
     & Serapion        &    &   &      &     &       &        \\
     & SystemDataMangt &    &      &      &     &      &       \\
     & TermoProjekt    &    &      &      &     &       &        \\
     & WorkFlow        &   &      &     &     &       &        \\ \bottomrule

\end{tabular}
 \begin{tablenotes}
    \small
          \item \checkmark the \textit{moderation} effect of size is significant.
           \item NA: metric data is not available. 
        \end{tablenotes}
     \end{threeparttable}
      \end{adjustbox}
      \label{tab:mod_individualSystems_summary_binary}
\end{table}

Based on our research hypothesis ({H02.1}), we cannot reject the null hypothesis that class size has no significant moderation effect on the relationship between defect-proneness and the following OO metrics:  RFC, WMC, CBO, LCOM and Fan-in. For Fan-out, the evidence of this effect is not consistent across the two datasets. We are unable to confirm if the moderation effect will always be present.

\subsection{A Comparison of the Systems Used}
\label{sec:similarityAnalysis}
In the previous results section, the concepts of moderation and mediation have been used to interpret the effects of class size and the validity of several OO metrics. All the mediation and moderation analysis results indicate that there is a considerable level of inconsistency between different datasets/systems. A quick look at the simple descriptive statistics (i.e., mean, variance, and correlation of various OO metrics, size and defects) indicates that data collected from different systems exhibit rather different patterns of instability, and that different systems tend to be distinct in their results. 

One way to compare any two different systems is by 
comparing the discrepancy between those systems through statistical hypotheses testing. Most statistical hypothesis tests are designed to accept/reject the pairwise equivalence of the underlying multivariate probability distributions. For example, it is possible to test if data obtained from two different systems follow the same multivariate distributions (a strict equivalence) or share the same correlation matrix (a weak equivalence). Although conducted, we did not report the results of these hypothesis tests here as both of them reject any possible equivalence between any two systems with extremely small p-values. 
A direct comparison between any two systems is also not practical given that those systems may contain different number of classes/files and highly-heterogeneous metrics values. The mean and variance of a single metric, size or number of defects in a system are not that informative in our case as we focus on the relationship between different variables. 
We are seeking a method to measure the similarity between any two systems while considering all metrics (i.e., size, OO metrics and defects).

A scale-free measure is needed to quantify the differences between systems. More specifically, the proposed metric should measure the distance of the inherent relationships of any two systems. 
Here, we make an effort to compare the systems in our dataset by constructing a suitable measure based on the correlation matrix and matrix norm. The correlation matrix of a system is a straightforward proxy of the internal relationships within the system. A matrix norm will induce a distance between two correlation matrices to compare the internal relationships of any two systems. 

For a system with $n$ classes, we collect the OO metrics and the number of defects in an $n\times p$ data matrix $X$. Here, $p=9$ variables includes the log size, log number of defects, and seven OO metrics. 
We used a scale-free correlation matrix, as the correlation coefficients are normalised and fall within a range between -1 and 1.
Let’s consider the data matrix $X$ of two systems $A$ and $B$ denoted by $X_A$ and $X_B$. The correlation matrices are $\Sigma_A=\left(\sigma_{ij}^{(A)}\right)$ and $\Sigma_B=\left(\sigma_{ij}^{(B)}\right)$, $1\le i,j\le p$ accordingly. To measure the difference between $\Sigma_A$ and $\Sigma_B$, we use the distance induced by the Frobenius norm as:

\begin{equation}
d(\Sigma_A,\Sigma_B)=\sqrt{\sum_{i,j}\left(\sigma_{ij}^{(A)}-\sigma_{ij}^{(B)}\right)^2}
\end{equation}

\begin{figure}[h]
\centering
\includegraphics[width=\linewidth]    {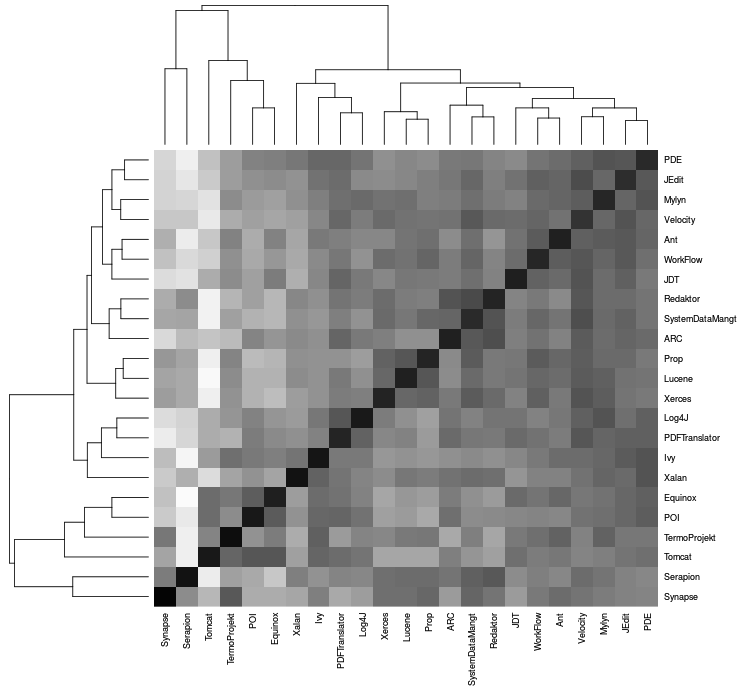}
    \caption{ Matrix heatmap for pairwise significance similarity measure with clustering}
     \label{fig:HeatMap_Similarity_Matrix}
\end{figure}

This distance is actually a pairwise measure which can be calculated for any two systems. The results are summarised in a distance matrix $D$ shown in the heatmap in Fig. \ref{fig:HeatMap_Similarity_Matrix}. Given the distance matrix $D$, we applied a hierarchical clustering algorithm to group similar systems based on the distance between each pair of systems, and then order them by similarity. We linked those clustering and correlation results with the results obtained from our mediation and moderation analysis, with the goal of investigating whether systems that are similar (i.e., in the same cluster group) have comparable results in terms of the mediation or moderation effect of size.

Table \ref{tab:dendogram_Matrix} provides a summary of the shared mediation and moderation effects of size on the metrics for systems in the same clusters (as shown in  Fig. \ref{fig:HeatMap_Similarity_Matrix}). In this table we focus on the top 12 clusters (i.e., clusters with the strongest correlations)
.
As shown in the dendrogram in Fig. \ref{fig:HeatMap_Similarity_Matrix} and the mediation analysis results in Tables \ref{tab:mediation_individualSystems_summary_conunt} and \ref{tab:mediation_individualSystems_summary_binary}, we observed that the majority of system pairs within the same cluster (high correlation) share the same mediation outcomes (with the majority of them either having no mediation effect 0f size at all, or the mediation effect appears for CBO only).
We also observed that systems that are clustered together and having strong correlation tend to share similar mediation more than moderation outcomes. For example, PDE and jEdit appear to be similar as they are in the same cluster and also have strong correlation. In both systems, size does not have any significant mediation effect on any of the OO metrics in both count or binary models. However, the results for the moderation effect of size for those two systems are dissimilar as the moderation effect of size appears only in RFC and LCOM (refer to Table \ref{tab:dendogram_Matrix}). This is also the case for Mylyn and Velocity, where both systems share a significant mediation effect of size for CBO and Fan-out in count models, and CBO in binary model, but also dissimilar in their moderation results.
However, for systems with lower correlation coefficients, the mediation and moderation results appear to be even more distinct. We conclude from this that systems with similar distance measures to some extent share the same mediation results for their individual metrics.

\begin{table}[]
\caption{Shared Mediation and Moderation effect of size on the OO metrics between systems in the same cluster}
  \begin{adjustbox}{max width=\textwidth}

\begin{tabular}{@{}llcc|cc@{}}
\toprule
                  &                   & \multicolumn{2}{c}{\textbf{Shared Mediation}} & \multicolumn{2}{c}{\textbf{Shared Moderation}} \\ \midrule
\textbf{System A} & \textbf{System B} & \textbf{Count}        & \textbf{Binary}       & \textbf{Count}            & \textbf{Binary}   \\ \midrule 
PDE               & jEdit             & No                    & No                    & RFC, LCOM                 & RFC               \\
Mylyn             & Velocity          & CBO, Fan-out          & CBO                   & No                        & No                \\
Serapion          & Synapse           & CBO                   & CBO                   & No                        & No                \\
PDE               & Mylyn             & No                    & No                    & RFC, LCOM                 & RFC               \\
Velocity          & jEdit             & No                    & No                    & No                        & No                \\
Ant               & Workflow          & No                    & No                    & No                        & No                \\
Redaktor          & SystemDataMangt   & No                    & No                    & No                        & No                \\
Prop              & Lucene            & No                    & CBO                   & WMC                       & No                \\
Log4J             & PDFTranslator     & No                    & No                    & No                        & No                \\
Equinox           & POI               & CBO                   & CBO                   & No                        & No                \\
Xerces            & Prop              & No                    & CBO                   & WMC                       & No                \\
Xerces            & Lucene            & \begin{tabular}[c]{@{}c@{}}CBO, Fan-in, \\ Fan-out\end{tabular} & \begin{tabular}[c]{@{}c@{}}CBO, Fan-in, \\ Fan-out\end{tabular} & \begin{tabular}[c]{@{}c@{}}RFC, WMC, CBO, \\ Fan-out\end{tabular} & LCOM, Fan-out   \\ 
\bottomrule
\end{tabular}
\end{adjustbox}
   \label{tab:dendogram_Matrix}

\end{table}

\subsection{Implications of the Results}
\label{sec:implications}
A number of prior studies have looked into the indirect effect of class size in defect models, with the assumption that the confounding effect of size is strong and impacts the performance of such models \citep{Benlarbi2001,Zhou2014}. The recommendation of those studies was that size should be controlled to avoid such an impact on the performance of the newly built models. 
Previous studies that looked into the effect of class size focused on binary defect models (i.e., fit a model to predict defect-proneness) and none looked into the impact of size in count models (i.e., fit a model to predict the number of defects). Our study has sought to address 1) the miscellany of the results from previous studies to confirm the indirect effect of size and 2) the effect of class size when looking at count models. We classified the indirect effect into a \textit{mediation} and \textit{moderation} effect, and we selected suitable statistical procedures to assess both.

The results presented here provide some evidence (although inconclusive) that size is in fact a problem for some OO metrics, and so should at least be considered. When looking into the OO metrics that we studied, the evidence regarding the impact of size is only evident for CBO (in both types of models, count and binary); and even then, it is shown in the majority of systems but not all of them. Other metrics show mixed results; and those results are far from conclusive. In fact, the results presented in this paper are somehow contradictory to those of the previous studies (i.e., \cite{Benlarbi2001,Zhou2014}). In terms of the moderation effect, another coupling metric, Fan-out, show a moderation effect of size in more systems that other metrics. Based on our in-depth investigation of the conditional (mediation) and interaction (moderation) effect of size, we cannot confirm if size is actually the reason for the relationship between most OO metrics and defects. 
We consider the effect of size to depend largely on the systems examined. For some systems, the effect of size affected all metrics, whereas for other systems there was no significant effect. This may be due to the nature of each specific system (i.e., type and domain of the system), or the dynamics of each development team (e.g., team experience, number of developers). We did not investigate these additional factors any further as such information are not always accessible for open-source projects. We provided an in-depth similarity analysis of the systems used in Section \ref{sec:similarityAnalysis}.

The main implication of these findings is that, rather than recommending that size should \emph{always} be controlled when building prediction models, we recommend that one should consider controlling for size only when significant evidence of mediation or moderation effects is observed, and that this  should be tested for each system individually.
Size remains an important metric for predicting defects (and other quality artefacts), but without proper investigation of the mediation and moderation effect of size for individual systems, size should not be omitted or controlled as a rule.

The results presented here provide a complementary perspective on the effect of class size, indicating that there is no strong evidence of a significant effect of size in defect prediction models. We note that the size effect depends on the systems examined. This is also the case for the moderation effect of size, where the findings regarding such effects are inconsistent across the systems studied. A further investigation of the nature of the systems used reveals that the data collected from different systems exhibit different patterns and therefore it is hard to generalise any results from this analysis. We expect the same outcomes from the previous studies as systems are quite different in nature. 
It does appear that the selection of datasets (the systems examined) plays an important role in determining whether size has a mediation or moderation effect on the association between OO metrics and defects.

We therefore recommend that if evidence of a significant mediation or moderation effect of size is found (which, as this paper advocates, should be examined using robust statistical methods as explained in Section \ref{sec:settings}), then one should consider controlling for size before looking at the prediction power of other OO metrics. Statistical procedures to control for the mediation or moderation effects are provided in the literature and should be followed if strong evidence regarding the indirect effect of size is found. 
There have been a number of methods proposed to control for  mediation or moderation effects, including a number of regression-based approaches \citep{dickinson2001finding,Zhou2014}. Stratification methods have also been suggested as an approach to control for mediators and other confounders \citep{mcnamee2005regression}.
In such cases (where there is evidence of an indirect effect of size), failing to control for size might result in models that are unstable and misleading in interpreting the prediction power of OO metrics. 

Our study fails in finding any strong and consistence evidence of a mediation or moderation effect of size across the systems that we investigated and for most metrics we included. Still, this does not mean that we should completely ignore (or even overestimate) the role of class size in such models. One may argue that size (LOC) can be used solely to predict the likelihood of defects given that size is correlated with the number of defects \citep{gyimothy2005empirical,herraiz2010beyond}. However, we believe that predicting defects using size alone can also be misleading. Other metrics should also be incorporated when such models are built. The mediation and moderation analysis can reveal the true impact of size on other metrics. 
However, class size can be used, with caution, as an initial indicator of the presence of defects in a class. LOC, for example, could be used as a good and quick predictor of defects, as it is generally easier to collect compared to other complexity metrics \citep{gyimothy2005empirical}.

The potential application of the \textit{mediation} and \textit{moderation} analysis presented in the paper goes beyond defects models, to other areas in software engineering. To the best of our knowledge, the concepts of \textit{mediation} and \textit{moderation} analysis have not been typically studied in software engineering research. The regression-based procedures we used in this study can potentially have a wider application in other software engineering  areas such as cost and effort estimation studies, software analytic studies, and social studies in software engineering (for example, in studies where there are human aspects with some potential indirect factors that may impact the outcomes, such as experience or familiarity with a technology).

\section{Threats to Validity}
\label{sec:threats}
This study has a number of validity threats.  We discuss possible threats to the validity of our study below.

The selected datasets pose a threat to our results. We selected data from two publicly available datasets. Most of the data collected are mainly from open-source Java-based projects (mainly Eclipse and Apache projects). However, we sought to select projects that have been used in previous defect prediction studies to allow us to compare our results with previous findings. Both datasets have been widely used in defect prediction studies in the past. In particular, the DAMB dataset has been used in the study of 
\cite{Zhou2014}. 
Also, we did not consider factors such as the domain of the program, and timeline of the development when we analysed our results. We believe that such information is important when building predication models. 

The method used for determining the mediation and moderation effects of size on defects prediction models can be a potential threat to the validity of the results. There are a number of statistical methods that are available to test for mediation and moderation effects. Previous studies have also utilised different methods to examine the confounding (indirect) effect of size. Those methods may show different conclusion due to their different power and robustness. We decided to use a bootstrapping-based CI methods following the recommendation in the literature \citep{MacKinnon2004ComparisonVariables,Hayes2013IntroductionApproach}, as those methods have been shown to be more robust compared to other available approaches.

\section{Conclusion}
\label{sec:conclusion}
Previous studies on defect-proneness have shown that class size has an indirect effect on the relationships between OO metrics and defect-proneness. However, no studies have attempted to study the same phenomenon in count defect prediction models. Our study reports on the analysis of the \textit{mediation} and \textit{moderation} effect of class size on the relationships between OO metrics and the number of defects/defect proneness. We investigated the potential indirect effect of size in count defect prediction models and re-examined the evidence on the indirect effect of size on defect-proneness.
When gauging indirect effects, bootstrapping-based techniques have several advantages over other methods, including the causal-steps and Sobel tests. 

By applying bootstrapping mediation we have shown that there is no conclusive evidence that size has a significant mediation effect on the relationships between most OO metrics and the number of defects, with evidence of a mediation effect of size on the relationship between CBO metric and defects. 
We also found inconsistent evidence that size has a moderation effect (i.e., affects the magnitude of the relationships between OO metrics and the number of defects), with a significant moderation effect found for RFC and Fan-out metrics. The findings for binary defect prediction models (defect proneness) are not different - our statistical analyses showed that there were inconsistent results regarding the mediation and moderation effects of class size on the relationship between individual OO metrics and defect proneness (again, only the CBO metric shows a significant results for mediation and the Fan-out metric for moderation). 

Unlike previous studies (e.g., \cite{Benlarbi2001,Zhou2014}), our results do not fully support the assertion that the size effect \textit{always exists} in count or binary models. In fact, our results suggest a contrary conclusion - that class size has no consistent significant mediation or moderation effect on the relationships between most OO metrics (all but CBO and Fan-out) and the number of defects or defect-proneness. 
We contend that bootstrapping-based techniques provide more powerful means for analysing mediator and moderator variables and we therefore recommend that empirical software engineering researchers employ such techniques when studying mediation and moderation effects. The approach we employed here can be applied to many problems in software engineering studies, and we encourage research to follow the steps we have employed here to study the the potential indirect effect of additional variables in their models.

\section{Acknowledgement}
The authors would like to thank the reviewers for the detailed and constructive comments on the earlier version of this paper, which were instrumental to improving the quality of the work.


\bibliographystyle{agsm}
\bibliography{main.bib}

@article{song2011general,
  title={A general software defect-proneness prediction framework},
  author={Song, Qinbao and Jia, Zihan and Shepperd, Martin and Ying, Shi and Liu, Jin},
  journal={IEEE Transactions on Software Engineering},
  volume={37},
  number={3},
  pages={356--370},
  year={2011},
  publisher={IEEE}
}

@inproceedings{Jureczko2010,
 author = {Jureczko, Marian and Madeyski, Lech},
 title = {Towards Identifying Software Project Clusters with Regard to Defect Prediction},
 booktitle = {Proceedings of the 6th International Conference on Predictive Models in Software Engineering},
 year = {2010},
 isbn = {978-1-4503-0404-7},
 location = {Timi\şoara, Romania},
 publisher = {ACM},
}

@article{bennin2017mahakil,
  title={MAHAKIL: Diversity based Oversampling Approach to Alleviate the Class Imbalance Issue in Software Defect Prediction},
  author={Bennin, Kwabena Ebo and Keung, Jacky and Phannachitta, Passakorn and Monden, Akito and Mensah, Solomon},
  journal={IEEE Transactions on Software Engineering},
  volume={44},
  number={6},
  year={2018},
  publisher={IEEE}
}

@article{he2012investigation,
  title={An investigation on the feasibility of cross-project defect prediction},
  author={He, Zhimin and Shu, Fengdi and Yang, Ye and Li, Mingshu and Wang, Qing},
  journal={Automated Software Engineering},
  volume={19},
  number={2},
  pages={167--199},
  year={2012},
  publisher={Springer}
}

@article{I,
    title = {{No Title}},
    author = {{\^{I}}, Ý and {\'{O}}, Ô and Ð, Ð Ý Þ and {\'{O}}, Ï and {\'{O}}, Ð and Þ, Ï and {\c{C}}, Ý Î Þ Ð and {\'{Y}}, Ð and {\^{U}}, Ï and {\'{I}}, Õ Þ}
}

@incollection{zimmermann2008predicting,
  title={Predicting bugs from history},
  author={Zimmermann, Thomas and Nagappan, Nachiappan and Zeller, Andreas},
  booktitle={Software Evolution},
  pages={69--88},
  year={2008},
  publisher={Springer}
}

@article{osman2018impact,
  title={The impact of feature selection on predicting the number of bugs},
  author={Osman, Haidar and Ghafari, Mohammad and Nierstrasz, Oscar},
  journal={arXiv preprint arXiv:1807.04486},
  year={2018}
}

@article{Ying2004,
 title={Predicting source code changes by mining change history},
  author={Ying, Annie TT and Murphy, Gail C and Ng, Raymond and Chu-Carroll, Mark C},
  journal={IEEE Transactions on Software Engineering},
  volume={30},
  number={9},
  year={2004},
  publisher={IEEE}
}

@article{Benlarbi2001,
    title = {{The Confounding Effect of Class Size on the Validity of Object-Oriented Metrics}},
    year = {2001},
    journal = {IEEE Transactions on Software Engineering},
    author = {El Emam, Kalhed and Benlarbi, Sa\"{\i}da and Goel, Nishith and Rai, Shesh N.},
    number = {7},
    pages = {630--650},
    volume = {27}
}

@article{Chidamber1994,
    title = {{A metrics suite for object oriented design}},
    year = {1994},
    journal = {IEEE Transactions on Software Engineering},
    author = {Chidamber, S.R. and Kemerer, C.F.},
    number = {6},
    month = {6},
    pages = {476--493},
    volume = {20},
    issn = {00985589}
}

@article{Zhou2014,
    title = {{An In-Depth Study of the Potentially Confounding Effect of Class Size in Fault Prediction}},
    year = {2014},
    journal = {ACM Transactions on Software Engineering and Methodology},
    author = {Zhou, Yuming and Xu, Baowen and Leung, Hareton and Chen, L I N},
    number = {1},
    volume = {23}
}

@article{Evanco2003Comments,
      title={Comments on "The Confounding Effect of Class Size on the Validity of Object-Oriented Metrics"},
  author={Evanco, William M},
  journal={IEEE Transactions on Software Engineering},
  volume={29},
  number={7},
  pages={670--672},
  year={2003},
  publisher={IEEE}
}

@article{Hayes2009BeyondMillennium,
    title = {{Beyond Baron and Kenny: Statistical Mediation Analysis in the New Millennium}},
    year = {2009},
    journal = {Communication Monographs},
    author = {Hayes, Andrew F.},
    number = {4},
    pages = {408--420},
    volume = {76},
    issn = {0363-7751}
}

@article{Preacher2008AsymptoticModels.,
    title = {{Asymptotic and Resampling Strategies for Assessing and Comparing Indirect Effects in Multiple Mediator Models}},
    year = {2008},
    journal = {Behavior Research Methods},
    author = {Preacher, Kristopher J and Hayes, Andrew F},
    number = {3},
    pages = {879--891},
    volume = {40},
}

@article{Basili1995,
  title={{a Validation of Object-Oriented Design Metrics As Quality Indicators}},
  author={Basili, Victor R and Briand, Lionel C. and Melo, Walc{\'e}lio L},
  journal={IEEE Transactions on Software Engineering},
  volume={22},
  number={10},
  pages={751--761},
  year={1996},
  publisher={IEEE}
}

@inproceedings{Zimmermann2007,
  title={Predicting Defects for Eclipse},
  author={Zimmermann, Thomas and Premraj, Rahul and Zeller, Andreas},
  booktitle={Proceedings of the 3rd International workshop on Predictor Models in Software Engineering},
  year={2007},
}

@article{Stinet1990,
    title = {{Direct and Indirect Effects : Classical and Bootstrap Estimates of Variability}},
    year = {1990},
    journal = {Sociological Methodology},
    author = {Bollen, Kenneth and Stinet, Robert},
    pages = {115--140},
    volume = {20}
}

@article{DAmbros2012EvaluatingComparison,
    title = {{Evaluating Defect Prediction Approaches: A Benchmark and an Extensive Comparison}},
    year = {2012},
    journal = {Empirical Software Engineering},
    author = {D'Ambros, Marco and Lanza, Michele and Robbes, Romain},
    number = {4-5},
    pages = {531--577},
    volume = {17},
    isbn = {1066401191739},
    issn = {13823256}
}

@article{Sobel1982AsymptoticModels,
    title = {{Asymptotic confidence intervals for indirect effects in structural equation models}},
    year = {1982},
    journal = {Sociological methodology},
    author = {Sobel, Michael E.},
    number = {1982},
    pages = {290--312},
    volume = {13},
}

@article{Hall2011AEngineering,
    title = {{A Systematic Review of Fault Prediction Performance in Software Engineering}},
    year = {2011},
    journal = {IEEE Transactions on Software Engineering},
    author = {Hall, Tracy and Beecham, Sarah and Bowes, David and Gray, David and Counsell, Steve},
    number = {6},
    pages = {1276--1304},
    volume = {38},
    isbn = {9781612081656},
    issn = {0098-5589}
}

@article{Hassan2009PredictingChanges,
    title = {{Predicting faults using the complexity of code changes}},
    year = {2009},
    journal = {Proceedings of the International Conference on Software Engineering},
    author = {Hassan, Ahmed E.},
    pages = {78--88},
}

@article{MacKinnon2002AEffects.,
    title = {{A Comparison of Methods to Test Mediation and Other Intervening Variable Effects}},
    year = {2002},
    journal = {Psychological Methods},
    author = {MacKinnon, David P and Lockwood, Chondra M and Hoffman, Jeanne M and West, Stephen G and Sheets, Virgil},
    number = {1},
    pages = {83--104},
    volume = {7},
}

@article{Fritz2007RequiredEffect,
    title = {{Required Sample Size to Detect the Mediated Effect}},
    year = {2007},
    journal = {Psychological Science},
    author = {Fritz, Matthew S and Mackinnon, David P},
    number = {3},
    pages = {233--239},
    volume = {18}
    }

@article{gyimothy2005empirical,
  title={Empirical Validation of Object-Oriented Metrics on Open Source Software for Fault Prediction},
  author={Gyimothy, Tibor and Ferenc, Rudolf and Siket, Istvan},
  journal={IEEE Transactions on Software Engineering},
  volume={31},
  number={10},
  pages={897--910},
  year={2005},
  publisher={IEEE}
}

@article{Basili1996HowSystems,
    title = {{How reuse influences productivity in object-oriented systems}},
    year = {1996},
    journal = {Commun. ACM},
    author = {Basili, Victor R. and Briand, Lionel C. and Melo, Walcélio L. Walc&#233;lio L},
    number = {10},
    pages = {104--116},
    volume = {39},
    issn = {0001-0782}
}

@article{Kitchenham2016RobustEngineering,
    title = {{Robust Statistical Methods for Empirical Software Engineering}},
    year = {2016},
    journal = {Empirical Software Engineering},
    author = {Kitchenham, Barbara and Madeyski, Lech and Budgen, David and Keung, Jacky and Brereton, Pearl and Charters, Stuart and Gibbs, Shirley and Pohthong, Amnart},
    number = {1},
    pages = {212--259},
    volume = {21},
    isbn = {9783642296444},
    issn = {15737616}
}

@article{Fenton1999AModels,
    title = {{A critique of Software Defect Prediction Models}},
    year = {1999},
    journal = {IEEE Transactions on Software Engineering},
    author = {Fenton, N E and Neil, M},
    number = {5},
    pages = {675--689},
    volume = {25},
    isbn = {0098-5589},
    issn = {00985589}
}

@article{Shepperd2014ResearcherPrediction,
    title = {{Researcher Bias: The Use of Machine Learning in Software Defect Prediction}},
    year = {2014},
    journal = {IEEE Transactions on Software Engineering},
    author = {Shepperd, Martin and Bowes, David and Hall, Tracy},
    number = {6},
    pages = {603--616},
    volume = {40},
    isbn = {013805326X},
    issn = {00985589},
    pmid = {1000106307}
}

@article{MacKinnon2004ComparisonVariables,
    title = {{Comparison of Approaches in Estimating Interaction and Quadratic Effects of Latent Variables}},
    year = {2004},
    journal = {Multivariate Behavioral Research},
    author = {MacKinnon, David P. and Lockwood, Chondra M. and Williams, Jason},
    number = {1},
    pages = {37--67},
    volume = {39},
    isbn = {0027-3171},
    issn = {0027-3171},
    pmid = {20157642}
}

@book{Hayes2013IntroductionApproach,
    title = {{Introduction to Mediation, Moderation, and Conditional Process Analysis: A Regression-based Approach}},
    year = {2013},
    author = {Hayes, Andrew F},
    publisher = {Guilford Press}
}

@article{Gil2017OnValidity,
   author = {Gil, Yossi and Lalouche, Gal},
 title = {On the Correlation Between Size and Metric Validity},
 journal = {Empirical Software Engineering},
 volume = {22},
 number = {5},
 year = {2017},
 issn = {1382-3256},
 pages = {2585--2611},
 publisher = {Kluwer Academic Publishers},
 address = {Hingham, MA, USA},
}

@article{baron1986moderator,
  title={The Moderator--Mediator Variable Distinction in Social Psychological Research: Conceptual, Strategic, and Statistical Considerations},
  author={Baron, Reuben M and Kenny, David A},
  journal={Journal of Personality and Social Psychology},
  volume={51},
  number={6},
  pages={1173},
  year={1986},
  publisher={American Psychological Association}
}

@inproceedings{schroter2006predicting,
  title={Predicting component failures at design time},
  author={Schr{\"o}ter, Adrian and Zimmermann, Thomas and Zeller, Andreas},
  booktitle={Proceedings of the International Symposium on Empirical Software Engineering},
  pages={18--27},
  year={2006},
  organization={ACM}
}

@inproceedings{tahir2018revisiting,
  title={Revisiting the size effect in software fault prediction models},
  author={Tahir, Amjed and Bennin, Kwabena E and MacDonell, Stephen G and Marsland, Stephen},
  booktitle={Proceedings of the 12th International Symposium on Empirical Software Engineering and Measurement},
  year={2018},
  organization={ACM}
}

@inproceedings{ghotra2015revisiting,
  title={Revisiting the impact of classification techniques on the performance of defect prediction models},
  author={Ghotra, Baljinder and McIntosh, Shane and Hassan, Ahmed E},
  booktitle={Proceedings of the 37th International Conference on Software Engineering},
  pages={789--800},
  year={2015},
  organization={IEEE}
}

@inproceedings{mende2009revisiting,
  title={Revisiting the evaluation of defect prediction models},
  author={Mende, Thilo and Koschke, Rainer},
  booktitle={Proceedings of the 5th International Conference on Predictor Models in Software Engineering},
  year={2009},
  organization={ACM}
}

@article{d2012evaluating,
  title={Evaluating defect prediction approaches: a benchmark and an extensive comparison},
  author={D’Ambros, Marco and Lanza, Michele and Robbes, Romain},
  journal={Empirical Software Engineering},
  volume={17},
  number={4-5},
  pages={531--577},
  year={2012},
  publisher={Springer}
}

@article{kamei2016studying,
  title={Studying just-in-time defect prediction using cross-project models},
  author={Kamei, Yasutaka and Fukushima, Takafumi and McIntosh, Shane and Yamashita, Kazuhiro and Ubayashi, Naoyasu and Hassan, Ahmed E},
  journal={Empirical Software Engineering},
  volume={21},
  number={5},
  pages={2072--2106},
  year={2016},
  publisher={Springer}
}

@article{pascarella2019fine,
  title={Fine-grained just-in-time defect prediction},
  author={Pascarella, Luca and Palomba, Fabio and Bacchelli, Alberto},
  journal={Journal of Systems and Software},
  volume={150},
  pages={22--36},
  year={2019},
  publisher={Elsevier}
}

@article{kamei2013large,
  title={A large-scale empirical study of just-in-time quality assurance},
  author={Kamei, Yasutaka and Shihab, Emad and Adams, Bram and Hassan, Ahmed E and Mockus, Audris and Sinha, Anand and Ubayashi, Naoyasu},
  journal={IEEE Transactions on Software Engineering},
  volume={39},
  number={6},
  pages={757--773},
  year={2013},
  publisher={IEEE}
}

@article{bennin2018relative,
  title={On the relative value of data resampling approaches for software defect prediction},
  author={Bennin, Kwabena Ebo and Keung, Jacky W and Monden, Akito},
  journal={Empirical Software Engineering},
  pages={1--35},
  volume={24},
  number={2},
  year={2018},
  publisher={Springer}
}

@article{catal2009investigating,
  title={Investigating the effect of dataset size, metrics sets, and feature selection techniques on software fault prediction problem},
  author={Catal, Cagatay and Diri, Banu},
  journal={Information Sciences},
  volume={179},
  number={8},
  pages={1040--1058},
  year={2009},
  publisher={Elsevier}
}

@article{turhan2009relative,
  title={On the relative value of cross-company and within-company data for defect prediction},
  author={Turhan, Burak and Menzies, Tim and Bener, Ay{\c{s}}e B and Di Stefano, Justin},
  journal={Empirical Software Engineering},
  volume={14},
  number={5},
  pages={540--578},
  year={2009},
  publisher={Springer}
}

@article{herraiz2010beyond,
  title={Beyond lines of code: Do we need more complexity metrics?},
  author={Herraiz, Israel and Hassan, Ahmed E},
  journal={Making software: what really works, and why we believe it},
  pages={125--141},
  year={2010},
  publisher={O’Reilly Media}
}

@article{tingley2014mediation,
  title={Mediation: R package for causal mediation analysis},
  author={Tingley, Dustin and Yamamoto, Teppei and Hirose, Kentaro and Keele, Luke and Imai, Kosuke},
  journal={Journal of Statistical Software},
  year={2014}, 
  volume = {59},
   number = {5}
}

@article{olague2007empirical,
  title={Empirical validation of three software metrics suites to predict fault-proneness of object-oriented classes developed using highly iterative or agile software development processes},
  author={Olague, Hector M and Etzkorn, Letha H and Gholston, Sampson and Quattlebaum, Stephen},
  journal={IEEE Transactions on software Engineering},
  volume={33},
  number={6},
  pages={402--419},
  year={2007},
  publisher={IEEE}
}

@article{zhou2006empirical,
  title={Empirical analysis of object-oriented design metrics for predicting high and low severity faults},
  author={Zhou, Yuming and Leung, Hareton},
  journal={IEEE Transactions on Software Engineering},
  volume={32},
  number={10},
  pages={771--789},
  year={2006},
  publisher={IEEE}
}

@inproceedings{dickinson2001finding,
  title={Finding failures by cluster analysis of execution profiles},
  author={Dickinson, William and Leon, David and Fodgurski, A},
  booktitle={Proceedings of the 23rd International Conference on Software Engineering},
  pages={339--348},
  year={2001},
  organization={IEEE}
}

@article{mcnamee2005regression,
  title={Regression modelling and other methods to control confounding},
  author={McNamee, Roseanne},
  journal={Occupational and Environmental Medicine},
  volume={62},
  number={7},
  pages={500--506},
  year={2005},
  publisher={BMJ Publishing Group Ltd}
}

@article{majumder2020revisiting,
  title={Revisiting Process versus Product Metrics: a Large Scale Analysis},
  author={Majumder, Suvodeep and Mody, Pranav and Menzies, Tim},
  journal={arXiv preprint arXiv:2008.09569},
  year={2020}
}

@inproceedings{binkley1998validation,
  title={Validation of the coupling dependency metric as a predictor of run-time failures and maintenance measures},
  author={Binkley, Aaron B and Schach, Stephen R},
  booktitle={Proceedings of the 20th International Conference on Software Engineering},
  year={1998},
  organization={IEEE}
}

@inproceedings{harrison1998coupling,
  title={Coupling metrics for object-oriented design},
  author={Harrison, Rachel and Counsell, Steve and Nithi, Reuben},
  booktitle={Proceedings of 5th International Software Metrics Symposium},
  year={1998},
  organization={IEEE}
}

@article{briand2000exploring,
  title={Exploring the relationships between design measures and software quality in object-oriented systems},
  author={Briand, Lionel C and W{\"u}st, J{\"u}rgen and Daly, John W and Porter, D Victor},
  journal={Journal of Systems and Software},
  volume={51},
  number={3},
  pages={245--273},
  year={2000},
  publisher={Elsevier}
}

@inproceedings{tang1999empirical,
  title={An empirical study on object-oriented metrics},
  author={Tang, Mei-Huei and Kao, Ming-Hung and Chen, Mei-Hwa},
  booktitle={Proceedings 6th International Software Metrics Symposium},
  year={1999},
  organization={IEEE}
}

@inproceedings{tantithamthavorn2018experience,
  title={An experience report on defect modelling in practice: Pitfalls and challenges},
  author={Tantithamthavorn, Chakkrit and Hassan, Ahmed E},
  booktitle={Proceedings of the 40th International Conference on Software Engineering: Software Engineering in Practice},
  year={2018},
    organization={IEEE}

}

@article{johnson2012rockchalk,
  title={rockchalk: Regression estimation and presentation},
  author={Johnson, Paul E and Grothendieck, G},
  journal={R package version},
  volume={1},
  number={2},
  year={2012}
}

@misc{repository,
  author = {Tahir, Amjed},
  title = {Does class size matter? online repository and replication package},
  year = {2021},
  journal = {GitHub repository},
  howpublished = {\url{https://github.com/amjedtahir/size-effect-defect-prediction}},
}

@inproceedings{zhang2009investigation,
  title={An investigation of the relationships between lines of code and defects},
  author={Zhang, Hongyu},
  booktitle={Proceedings of the International Conference on Software Maintenance},
  year={2009},
  organization={IEEE}
}

@article{hayakawa2021novel,
  title={A novel approach to address external validity issues in fault prediction using bandit algorithms},
  author={Hayakawa, Teruki and Tsunoda, Masateru and Toda, Koji and Nakasai, Keitaro and Tahir, Amjed and Bennin, Kwabena Ebo and Monden, Akito and Matsumoto, Kenichi},
  journal={IEICE TRANSACTIONS on Information and Systems},
  volume={104},
  number={2},
  pages={327--331},
  year={2021},
  publisher={IEICE}
}

\end{document}